\documentclass[a4paper,11pt]{article}
\usepackage{units}
\usepackage{float}
\usepackage{amsmath}
\usepackage{amssymb}
\usepackage{graphicx}
\usepackage{booktabs}
\usepackage{caption}
\usepackage{subcaption}
\usepackage{multirow}
\usepackage{algorithm}
\usepackage[noend]{algpseudocode}
\usepackage[section]{placeins}

\pdfoutput=1

\usepackage{jheppub}
\title{\boldmath The twisted gradient flow coupling at one loop}

\author{Eduardo I. Bribi\'an, }
\author{Margarita Garc\'{\i}a P\'erez}
\affiliation{Instituto de F\'{\i}sica Te\'orica UAM-CSIC,\\
       C/ Nicol\'as Cabrera 13-15,
       Universidad Aut\'onoma de Madrid,\\
       28049 Madrid, Spain\\
       }

\emailAdd{e.i.bribian@csic.es}
\emailAdd{margarita.garcia@uam.es}

\abstract{

We compute the one-loop running of the $SU(N)$ 't Hooft coupling in a finite volume gradient flow scheme using twisted boundary conditions. The coupling is defined in terms of the energy density of the gradient flow fields at a scale $\tl$ given by an adequate combination of the torus size and the rank of the gauge group, and is computed in the continuum using dimensional regularization. We present the strategy to regulate the divergences for a generic twist tensor, and determine the matching to the $\MS$ scheme at one-loop order. For the particular case in which the twist tensor is non-trivial in a single plane, we evaluate the matching coefficient numerically and determine the ratio of $\Lambda$ parameters between the two schemes. We analyze the $N$ dependence of the results and the possible implications for non-commutative gauge theories and volume independence.
}


\preprint{%
{\flushright
IFT-UAM/CSIC-19-26
\\
}}

\newcommand{\comment}[1]{}

\newcommand{\be}{\begin{equation}}
\newcommand{\ee}{\end{equation}}
\newcommand{\ba}{\begin{array}}
\newcommand{\ea}{\end{array}}
\newcommand{\baa}{\begin{array}}
\newcommand{\eaa}{\end{array}}
\newcommand{\bea}{\begin{eqnarray}}
\newcommand{\eea}{\end{eqnarray}}
\newcommand{\bal}{\begin{align}}
\newcommand{\eeal}{\end{align}}
\newcommand{\half}{\frac{1}{2}}

\newcommand{\Tr}{\mathrm{Tr}}

\newcommand{\hc}{{\hat c}}
\newcommand{\tl}{\tilde l}

\newcommand{\htheta}{\hat \theta}
\newcommand{\TGF}{{\rm TGF}}
\newcommand{\cE}{{\mathcal E}}

\newcommand{\wtsum}{\underset{q,r}{\widehat{\sum}}}
\newcommand{\ts}{\tilde s}
\newcommand{\tu}{\tilde u}
\newcommand{\tv}{\tilde v}
\newcommand{\rd}{{\rm div}}
\newcommand{\rf}{{\rm fin}}

\newcommand{\MS}{{\overline{\rm MS}}}
\newcommand{\cA}{{\mathcal A}}
\newcommand{\cN}{{\mathcal N}}
\newcommand{\p}{p}
\newcommand{\q}{q}
\newcommand{\B}{\hat B}
\newcommand{\A}{\hat A}
\newcommand{\cV}{\mathcal{V}}
\newcommand{\Z}{ \mathbb{Z}}
\newcommand{\cI}{ \mathcal{I}}
\newcommand{\cC}{ \mathcal{C}}
\newcommand{\Phio}{ \Phi^{(0)}}

\newcommand{\kb}{ \bar k}
\newcommand{\tI}{ \bar I}

\begin{document}

\maketitle
\flushbottom

\newpage
\section{Introduction}
In recent years, the continuous smoothing procedure known as the gradient flow ~\cite{Narayanan:2006rf,Luscher:2009eq,Luscher:2010iy} has received considerable attention. One of its most common applications has been, in combination with finite-size scaling techniques, the determination of the non-perturbative scale dependence of the gauge coupling constant. Examples of the usefulness of this approach range from precise, non-perturbative determinations of the QCD coupling constant and $\Lambda$ parameter~\cite{Brida:2014joa,DallaBrida:2016kgh,Bruno:2017gxd,Korzec:2017ypb,Ishikawa:2017xam} to the study of Yang-Mills theories with near conformal behavior~\cite{Lin:2015zpa,Fodor:2016zil,Hasenfratz:2016dou}, or with large number of colors~\cite{GarciaPerez2015}. Several coupling renormalization schemes based on gradient flow techniques have been proposed to that end~\cite{Luscher:2010iy,Fodor:2012td,Borsanyi2012,Fritzsch:2013je,Ramos:2014kla}. These schemes can be related through a perturbative calculation to more traditional ones such as the $\MS$ scheme, a step often required to make contact with experiment. However, and despite their importance, perturbative calculations in such a set-up are scarce. In infinite volume, the matching has been determined up to both next-to-leading order (NLO)~\cite{Luscher:2010iy} and next-to-next-to-leading order (NNLO)~\cite{Harlander:2016vzb}, while for finite volume it has been done up to NNLO in the Schr\"odinger functional scheme using numerical stochastic perturbation theory~\cite{DallaBrida:2016dai,DallaBrida:2017tru}.  

In the scope of this work, we will focus on a particular gradient flow finite volume scheme for $SU(N)$ pure gauge theory, introduced by A.Ramos in ref.~\cite{Ramos:2014kla}. We will be presenting results for the matching at NLO of this scheme to the $\MS$ one, determining the coupling scale in terms of the size of a 4-dimensional torus endowed with twisted boundary conditions (TBC)~\cite{tHooft:1979rtg}. From the point of view of perturbative calculations, TBC have an enormous advantage over periodic ones (PBC), as using TBC turns the set of zero-action solutions into a discrete one, and avoids the quartic nature of the fluctuations around $A_\mu=0$ present with PBC~\cite{GonzalezArroyo:1981vw}.
The usefulness of TBC for perturbation theory was first formulated in the context of volume reduction in large $N$ Yang-Mills theory~\cite{GonzalezArroyo:1982ub,GonzalezArroyo:1982hz}, and was then extended to various other contexts at finite and large $N$~\cite{Fabricius:1985jw,Luscher:1985zq,Luscher:1985wf,Coste:1986cb,Hansson:1986ia,GonzalezArroyo:1988dz,Daniel:1989kj,Daniel:1990iz,Snippe:1996bk,Snippe:1997ru,Perez:2013dra,Gonzalez-Arroyo:2014dua,Perez:2017jyq,Perez:2018afi,Gonzalez-Arroyo:2019zfm}. Despite these advantanges, as we will show along this work, perturbative calculations in the twisted gradient flow scheme remain challenging, although so far an analogous perturbative calculation in the case of periodic boundary conditions has not been obtained.

Our interest in this calculation goes beyond the particular applicability of the results, and connects with theoretical ideas related to the concept of volume independence in gauge theories. An essential ingredient of the construction has to do with the dependence of the coupling on the number of colors. We will follow the finite-volume prescription adopted in~\cite{GarciaPerez2015}, and set the $SU(N)$ running coupling scale to be proportional to an effective size $\tl$ combining the torus size and the number of colors. In perturbation theory, this effective scale is expected to jointly capture the dependence of the coupling on the volume and on $N$, once an angular variable depending on the choice of twist is fixed. 
For the purposes of this paper, we will keep $N$ finite but use our results to analyze the dependence of the coupling on $N$, at large values of $N$. Setting the energy scale of the renormalized 't Hooft coupling to be $\mu = 1/(c \tl)$, we will consider two different types of $N\rightarrow \infty$ limits, a {\it thermodynamic} limit in which the effective size is sent to infinity as $c$ goes to zero while $\mu$ is kept constant, and a second one, called singular in \cite{AlvarezGaume:2001tv}, in which $N$ is sent to infinity as the torus size is shrunk to zero at constant $c$ in such a way as to keep $\tl$ fixed~\cite{Perez:2013dra,GarciaPerez2015,Chamizo:2016msz,Perez:2018afi}~\footnote{This type of limit has also been considered in other works such as~\cite{Guralnik:2001pv,Unsal:2008ch,Unsal:2010qh}. See~\cite{Unsal:2018} and references therein as well for more examples.}.
In ref.~\cite{GarciaPerez2015} this last limit was used to compute the $SU(\infty)$ running coupling through a step scaling procedure in which the step size was modified via changes in the rank of the gauge group.

This particular prescription for scale setting was inspired by the idea of volume reduction in lattice gauge theories. Originally formulated by Eguchi and Kawai~\cite{Eguchi:1982nm}, volume reduction states that in the ({\it thermodynamic}) large $N$ limit, $SU(N)$ theory becomes independent of the physical size of the torus.  
Proof for this statement relies on the independence of the large $N$ Schwinger-Dyson equations from lattice volume, which in turn requires center symmetry to be preserved. As the symmetry was shown not to hold with PBC~\cite{Bhanot:1982sh}, several alternative proposals were formulated~\cite{Bhanot:1982sh,Gross:1982at,Narayanan:2003fc,Kovtun:2007py,Unsal:2008ch}, one of which was the use of twisted boundary conditions~\cite{GonzalezArroyo:1982hz,GonzalezArroyo:1982ub}, which has proven very successful provided the twist tensor is judiciously chosen~\cite{GonzalezArroyo:2010ss,GonzalezArroyo:2012fx,Gonzalez-Arroyo:2013bta,Gonzalez-Arroyo:2014dua,Perez:2015yna,Gonzalez-Arroyo:2015bya,Perez:2017jyq,Gonzalez-Arroyo:2018aus}.
The idea of volume reduction with TBC was extended to the continuum theory in ref.~\cite{GonzalezArroyo:1983ac}, constituting the first formulation of the Feynman rules of a non-commutative Yang-Mills theory~\cite{Douglas:2001ba}. The singular limit in ref.~\cite{AlvarezGaume:2001tv} was first formulated precisely in the context of such non-commutative theories, with the effective torus size $\tl$ arising in a natural way through the Morita duality. Thus, ordinary gauge theories on a twisted torus are related to non-commutative theories with a rational value of the dimensionless non-commutativity parameter, the effective size corresponding to the one of the non-commutative torus. In 2+1 dimensions, several recent works~\cite{Chamizo:2016msz,Perez:2018afi} have analyzed the possibility of defining non-commutative gauge theories at irrational values of the non-commutativity parameter as the limit of a sequence of ordinary twisted gauge theories of an increasing number of colors. These works have shown that, if one wishes to avoid tachyonic instabilities~\cite{Guralnik:2002ru}, such a construction can only be achieved for an uncountable, zero-measure set of values of the non-commutativity parameter. We will, in light of these results, analyze the behavior of the coupling in the singular large $N$ limit.

In this paper, we present a perturbative calculation in the continuum of the running `t Hooft coupling constant at NLO. The layout of the paper is as follows: in sec.~\ref{s:tgfcoupling}, we introduced the twisted gradient flow (TGF) scheme, presenting the gradient flow observable used to define the running coupling (i.e. the energy density evaluated at a positive flow time proportional to an effective size $\tl$), along with some specifics about the implementation of TBC in our setting. In particular, we detailed the orthogonal twist used throughout the paper. Sec.~\ref{s:perturbative} then presents the perturbative expansion, along with the regularization and renormalization schemes. This is the longest, most technical section, and contains a fair bit of algebraic manipulation. The calculation is analogous to the one performed by L\"uscher in infinite volume~\cite{Luscher:2010iy}, though many particularities to the twisted finite volume scheme appear. We will simply mention that it contains the expansion of the observable in powers of the coupling, a reformulation of the NLO contribution as the sum of several integrals, the identification of the divergent terms entering the calculation, and a procedure to regularize them by relating them to infinite volume expressions that can be evaluated in dimensional regularization. Expressions for the observable at LO and NLO are provided within the section, but any reader interested in the final expression for the matching at one-loop order of the TGF coupling to the $\MS$ scheme may skip directly to sec.~\ref{s:oneloop}, which contains both the matching to the $\MS$ scheme and the ratio of $\Lambda$ parameters (which need to be computed numerically). Results for the case of a 2-dimensional non trivial twist and several $SU(N)$ groups are presented in subsection~\ref{s:results}. Sec.~\ref{s:colours} discusses the dependence of the coupling on the number of colors, following similar arguments to those in~\cite{Perez:2018afi,Perez:2017jyq}. A summary of results is presented in sec.~\ref{s:conclusions}. Many technicalities were moved for clarity to appendices~\ref{ap:frules}-~\ref{ap:numerical}, including details on the algorithms used to compute the $\Lambda$ parameter.

\section{The twisted gradient flow coupling}
\label{s:tgfcoupling}

One of the applications of the gradient flow method has been the computation of the Yang-Mills running coupling, using the energy density $E(t)$ of the gradient flow field as the defining observable. At positive, non-zero flow time $t$, $\langle t^2 E(t) \rangle$ is a renormalized quantity and, at leading order in perturbation theory, is proportional to the $\MS$ coupling at a scale $\mu = 1/\sqrt{8t}$, which leads to a natural definition of a renormalized coupling constant~\cite{Luscher:2010iy}. In this work, we focus on a particular gradient flow scheme that makes use of finite size scaling on a torus with TBC. As discussed in the introduction, our set-up is based on the one introduced by A. Ramos in ref.~\cite{Ramos:2014kla}, but differs slightly from it for reasons that will become clear in what follows.

\subsection{The definition of the coupling}
\label{s:tgfdef}

The gradient flow is based on the introduction of a parameter $t$, known as flow time, in such a way as to define a $t$-dependent gauge field $B_{\mu}\left(x,t\right)$ matching the Yang-Mills one $A_{\mu}\left(x\right)$ at $t=0$.  As flow time passes, this gauge field is smeared down towards the minimum action solutions, its evolution driven by the so-called flow equations:
\begin{equation}
\partial_{t}B_{\mu}\left(x,t\right)=D_{\nu}G_{\nu\mu}\left(x,t\right),\qquad B_{\mu}\left(x,0\right)=A_{\mu}\left(x\right),
\end{equation}
where $D_{\mu}$ and $G_{\mu\nu}$ respectively stand for the covariant derivative and field strength tensor of the flow fields:
\begin{align}
D_{\mu}B_{\nu}\left(x,t\right)&=\partial_{\mu}B_{\nu}\left(x,t\right)+i\left[B_{\mu}\left(x,t\right),B_{\nu}\left(x,t\right)\right],\\
G_{\mu\nu}\left(x,t\right)&=\partial_{\mu}B_{\nu}\left(x,t\right)-\partial_{\nu}B_{\mu}\left(x,t\right)+i\left[B_{\mu}\left(x,t\right),B_{\nu}\left(x,t\right)\right].
\end{align}
This scheme is particularly useful, as observables built from the expectation values of products of $B$ fields at positive flow time have been shown to be renormalized quantities \cite{Luscher:2011bx}.

The renormalized gradient flow coupling can then be defined in terms of the energy density of the flowed field:
\begin{equation}
\langle E(t) \rangle = \half \Big\langle \Tr \Big( G_{\mu\nu}(x,t) G_{\mu\nu}(x,t) \Big )\Big\rangle.
\label{eq:edens}
\end{equation}
In infinite volume, this quantity can be used to define a renormalized $SU(N)$ 't Hooft coupling at an energy scale $\mu$, given by~\cite{Luscher:2010iy}:
\begin{equation}
\lambda(\mu )= \left .\frac{128 \pi^2 N^2 }{(d-1) (N^2-1)} \, \left\langle \frac{ t^2 E(t) }{N}\right\rangle\right|_{t=1/(8\mu^2)} ,
\end{equation}
where $d$ stands for the number of space-time dimensions of the theory. Finite volume gradient flow schemes~\cite{Fodor:2012td,Borsanyi2012,Fritzsch:2013je,Ramos:2014kla,GarciaPerez2015} use a formulation 
in which the gauge theory is defined on a finite torus instead, with each scheme differing in specific details such as, for instance, different boundary conditions. The most common choice is to use a symmetric torus, with all directions of equal length $l$, while setting the scale for the renormalized coupling in terms of $l$ by fixing $\mu= 1/(cl)$, with $c$ an arbitrary constant. Each specific choice of $c$, always taken to be smaller than unity, is an intrinsic part of the definition of the scheme.

The $SU(N)$ TGF coupling used in this paper is inspired in the finite-volume schemes proposed in refs.~\cite{Ramos:2014kla,GarciaPerez2015,Keegan:2015lva}. We will leave the specifics of the scheme for the next subsection, but mention that our $SU(N)$ gauge theory will be defined on an 4-torus with TBC~\cite{tHooft:1979rtg}, and such that the torus has a period $l$ in $d_t$ (twisted) directions, and $\tl= l N^{2/d_t}$ in the remaining $4-d_t$ ones, with $d_t$ being either two or four. The reasons behind our choice of an asymmetric torus will become clear in what follows. In this scheme, the twisted gradient flow 't Hooft coupling is defined by~\cite{GarciaPerez2015}:  
\begin{equation}
\lambda_\TGF(c\tl)=\left.\mathcal{F}(c) \, \left\langle \frac{ t^2 E(t) }{N}\right\rangle \right|_{t=c^2\tl^2/8},
\label{eq:ltgf}
\end{equation}
where $\mathcal{F}(c)$ is a constant defined in such a way as to have $\lambda_\TGF(\tilde{l}) = \lambda_0 + {\mathcal O} (\lambda_0^2)$, in terms of the bare 't Hooft coupling $\lambda_0$.

\subsection{The choice of boundary conditions and torus size}
\label{s:choices}

In this subsection, we will discuss the particular definition of the TGF scheme used in this paper. The main idea is to have a perturbative set-up that is as symmetric as possible~\cite{GarciaPerez2015,Keegan:2015lva}. To achieve this, we will look at the quantization of momenta in our particular setting, and select the torus size accordingly. We will begin with a generic discussion of the quantization of momenta in the presence of TBC, leading to the introduction of $\tl$ as the relevant length scale. 
 
Let us start by considering a $SU(N)$ gauge theory defined on a $d$-dimensional torus of length $l_\mu$ in each direction, and focus our analysis in the specific case of four dimensions for a gauge potential that satisfies 't Hooft TBC~\cite{tHooft:1979rtg}. We will work with an orthogonal twist, for which the gauge potential can be fixed to be periodic in each direction up to a constant gauge transformation:
\be
A_\mu ( x + l_\nu \hat \nu) = \Gamma_\nu A_\mu (x) \Gamma_\nu^\dagger \, ,
\label{eq:tbc0}
\ee 
where $ \Gamma_\nu$ are four $SU(N)$ matrices known as twist eaters which satisfy:
\be
\Gamma_\mu \Gamma_\nu  = Z_{\mu \nu} \Gamma_\nu \Gamma_\mu ,
\label{eq:tweat}
\ee
with $Z_{\mu \nu}$ an element of the center of the gauge group, written in terms of an antisymmetric tensor of integers $n_{\mu \nu}$ as:
\be
 Z_{\mu \nu} = \exp\{ i 2 \pi n_{\mu \nu} /N\}\, .
\ee
The twist tensor $n_{\mu \nu}$ is preserved under gauge transformations, and uniquely characterizes the boundary conditions. It is said to be orthogonal when $\kappa(n)= \epsilon_{\mu \nu \rho\sigma} n_{\mu \nu} n_{\rho \sigma}/8 = 0$ (mod $N$). Among such tensors, we will focus only in the so-called irreducible twist tensors, which are the subset for which the only matrices that commute with all $\Gamma_\mu$ are the ones proportional to the identity in $SU(N)$.

Irreducible twist tensors have been known to be advantageous for perturbative calculations, as the class of gauge-inequivalent zero-action solutions is discrete~\cite{tHooft:1981nnx,vanBaal:1983eq} and zero-modes are eliminated, making computations in perturbation theory much easier. A detailed discussion of the conditions under which a twist is irreducible can be found in \cite{ga:torus}. For the scope of this work, we will focus on two types of irreducible twist tensors (detailed below), which are non-trivial in either a single plane or in all of them. For the sake of clarity in the description, we will use gauge freedom to impose strict periodicity for the gauge potential in all directions except for a number $d_t$ of them, dubbed "twisted directions", taken to be either two or four, though the specific form of the twist matrices is irrelevant as long as  eq.~\eqref{eq:tweat} is satisfied. We will write our orthogonal twist tensor in the form:
\be
 n_{\mu \nu} = \epsilon_{\mu \nu}  \frac{k N}{l_g}\, ,
\label{eq:twtensor}
\ee
where $l_g = N^{2/d_t}$ depends on both the number of colors and the number of twisted directions, and $k$ and $l_g$ are two coprime integers that guarantee that the irreducibility condition is satisfied. The choice to have a non-trivial twist in only the (0,1) plane is made by setting $d_t=2$ and $\epsilon_{0 1}=-\epsilon_{1 0}=1$, and by choosing $\epsilon_{\mu \nu} =0$ in any other plane, whereas to twist all planes non-trivially one must instead take $d_t=4$, and set $\epsilon_{\mu \nu}$ to be antisymmetric and equal to 1 whenever $\mu < \nu$.
With this choice:
\be
 Z_{\mu \nu} = \exp\left\{ i 2 \pi \epsilon_{\mu \nu}  \frac{k}{l_g}\right \}\, .
\label{eq:twist}
\ee

A non-trivial twist, such as the one above, will affect the quantization of momenta in the finite box.

The solution to the boundary conditions on such twisted tori in the continuum is well known~\cite{GonzalezArroyo:1982hz}, as one can see for instance in~\cite{Perez:2017jyq} for the general treatment when the torus is discretized on a lattice, or in~\cite{Perez:2013dra} for an example in 2+1 dimensions in continuum perturbation theory. We will, in what is left of this subsection, recall some known results necessary to implement perturbation theory with TBC. We start by defining:
\be
\hat{\Gamma}(q)=\frac{1}{\sqrt{2N}}e^{i\beta\left(q\right)}\Gamma_{0}^{s_{0}\left(q\right)}\dots\Gamma_{d_t-1}^{s_{d_t-1}\left(q\right)}\, ,
\ee
with $s_\mu(q) \in \mathbb{Z}$.
Provided $k$ and $l_g$ are coprime integers, there will be $N^2$ independent $SU(N)$ matrices of this type, of which the only non-traceless one is the one proportional to the identity matrix, i.e. the one for which $s_\mu(q) = 0\, ({\rm mod}\, l_g)$ in all twisted directions. Excluding it, the remaining $N^2-1$ matrices constitute a basis for the $SU(N)$ Lie algebra.  
If $\hat{\Gamma}(q)$ satisfies:
\be
\Gamma_\nu  \hat{\Gamma}(q) \Gamma_\nu^\dagger = e^{i q_\nu l_\nu}  \hat{\Gamma}(q)\, ,
\label{eq:tbcg}
\ee
with no summation over $\nu$ implied, the boundary conditions in \eqref{eq:tbc0} are trivially implemented through the Fourier expansion:
\be
A_\mu ( x) = V^{-\half} \underset{q}{\sum^{\prime}} e^{i q x} \hat A_\mu (q) \, \hat{\Gamma}(q)\, ,
\label{eq:fourier}
\ee
where $V\equiv \prod_\mu l_\mu$, and the prime in the sum denotes the exclusion of the momenta for which $\hat{\Gamma}(q) \propto \mathbb{I}$. In the periodic directions, for which $\Gamma_\nu\propto \mathbb{I}$, the momenta are as usual quantized in units of $2\pi/l_\nu$. This is however not the case for the twisted directions, where a solution is provided by:
\be
s_{\mu}(q)=\tilde{\epsilon}_{\mu\nu}  \bar k \, \frac{l_g l_\nu q_{\nu}}{2\pi}  \in \mathbb{Z},
\,
\ee
where $\bar{k}$ and $\tilde{\epsilon}_{\mu\nu}$ are given by:
\be
k\bar{k}=1 \,  ({\rm mod}\, l_{g}),\qquad\underset{\nu}{\sum}\tilde{\epsilon}_{\mu\nu}\epsilon_{\nu\rho}=\delta_{\mu\rho}.
\ee
The momentum along the twisted $\mu$ directions is thus quantized in units of $2\pi/\tl_\mu$, with $\tl_\mu \equiv l_\mu l_g$. For this choice of $s_\mu$, the group structure constants in the $\hat{\Gamma}(q)$ basis become momentum dependent and are given by:
\be
\left[\hat{\Gamma}\left(p\right),\hat{\Gamma}\left(q\right)\right]=iF\left(p,q,-p-q\right)\hat{\Gamma}\left(p+q\right)\, ,   
\label{eq:comm} 
\ee
with
\be
F\left(p,q,-p-q\right)=-\sqrt{\frac{2}{N}}\sin\left(\frac{1}{2}\theta_{\mu\nu}p_{\mu}q_{\nu}\right) \, ,
\label{eq:Fdef}
\ee
and
\be
\theta_{\mu\nu}=\frac{\tl_\mu \tl_\nu}{2\pi} \, \tilde \epsilon_{\mu \nu} \htheta,\qquad   \htheta  = \frac{ \bar k}{l_g} .
\ee
The tracelessness of the $\hat{\Gamma}(q)$ matrices thus forbids momenta such that $\tl_\mu q_\mu = 0 \, ({\rm mod}\, 2 \pi l_g)$ in all twisted directions, and so in particular it forbids zero momentum in the twisted box. 

The previous analysis implies that momentum is quantized differently in periodic and twisted directions: it is quantized in terms of the inverse torus size for the former, and in terms of an effective size combining the torus period and the number of colors of the gauge group, $\tl_\mu= l_\mu l_g = l_\mu N^{2/d_t}$, for the latter. This observation has led us to a specific choice of torus size to define the TGF coupling in eq.~\eqref{eq:ltgf}, picked in such a way as to impose the same momentum quantization in all directions. When $d_t=2$, this will be achieved by considering an asymmetric torus of length $l$ in the twisted directions and $\tl$ in the periodic ones, whereas for $d_t=4$ we will
instead pick a symmetric 4-torus of period $l$ in all directions. This way, all momenta will always be quantized in units of $2\pi / \tl$, and we will use this effective size $\tl$ as the renormalization scale for the running coupling.

\section{Perturbative expansion}
\label{s:perturbative}

The procedure to determine the perturbative expansion of the coupling follows closely the one developed by L\"uscher in infinite volume in ref.~\cite{Luscher:2010iy}. The main difference arises from the quantization of momentum on the torus, as momentum integrals become sums over an infinite set of discrete momenta, and from the change in the group structure constants due to the different choice of $SU(N)$ Lie algebra basis. Divergent momentum sums, however, can still be treated via dimensional regularization -- see for instance \cite{Brezin:1985xx} -- in a way that will be detailed in this section.

\subsection{Perturbative expansion of the energy density}

As a first step towards obtaining the perturbative expansion of the observable, we will fix the gauge in such a way that the following periodicity conditions are satisfied:
\begin{align}
A_\mu(x + l \hat \nu) &= \Gamma_\nu A_\mu(x) \Gamma_\nu^\dagger, \qquad   \text{ for} \  \nu = 0, \cdots, d_t-1 , \label{eq:tbc} \\
A_\mu(x + \tl \hat \nu) &= A_\mu(x) , \qquad   \text{ for} \  \nu = d_t, \cdots, d-1 ,
\label{eq:pbc}
\end{align}
where $\Gamma_\mu$ satisfies eq.~\eqref{eq:tweat}, and with a twist tensor of the form shown in eq.~\eqref{eq:twist}. This restricts the set of allowed gauge transformations $\Omega(x)$ down to those preserving the form of the twist matrices, i.e. those satisfying:
\begin{align}
\Omega(x + l \hat \nu) &= \Gamma_\nu \Omega(x) \Gamma_\nu^\dagger, \qquad   \text{ for} \  \nu = 0, \cdots, d_t-1 , \\
\Omega(x + \tl \hat \nu) &= \Omega(x) , \qquad   \text{ for} \  \nu = d_t, \cdots, d-1 .
\label{eq:gtbc}
\end{align}
These boundary conditions are implemented through the Fourier expansion of the gauge field given in eq.~\eqref{eq:fourier}. In the specific case of the asymmetric torus that we are considering, the torus volume is given by $V = l^{d_t} \tl^{\, d-d_t}$, and momenta in all directions are quantized in terms of the effective size $\tl$. As we recall, the prime in the sum in eq.~\eqref{eq:fourier} denotes the exclusion of all momenta for which $\tl q_\mu = 0$  (mod  $2\pi l_g)$ in all twisted directions, which in particular excludes zero modes.

With this, we may begin the perturbative expansion, which we perform around the $A_\mu=0$, zero-action solution. We start in $d=4-2 \epsilon$ dimensions by scaling the original gauge potential with the bare coupling, $A_\mu(x) \rightarrow g_0 A_\mu(x)$. The full Feynman rules in momentum space, given in the Feynman gauge and derived using the boundary condition-preserving Fourier representation mentioned in the previous section, can be found in appendix~\ref{ap:frules}.

It will be convenient to henceforth use a set of modified flow equations:
\begin{equation}
\partial_{t}B_{\mu}\left(x,t\right)=D_{\nu}G_{\nu\mu}\left(x,t\right)+\xi D_{\mu}\partial_{\nu}B_{\nu}\left(x,t\right),\quad B_{\mu}\left(x,0\right)=g_0 A_{\mu}\left(x\right),
\label{eq:modflow}
\end{equation}
$\xi$ being a gauge parameter to be set to unity. At fixed $t$, the field derived from this modified flow equation can be related to the solution of the original one by a gauge transformation~\cite{Luscher:2010iy}, and hence the modification does not affect gauge invariant observables such as the one we are considering in this paper. It can be shown that the corresponding (flow-time dependent) gauge transformation preserves the boundary conditions \eqref{eq:tbc} at any given flow time~\cite{Ramos:2014kla}. These modified flow equations can be solved order by order in $g_0$ by expanding the flow field in powers of the coupling:
\begin{equation}
B_{\mu}(x,t)=\underset{k}{\sum} \, g_{0}^{k}B_{\mu}^{(k)}(x,t),\quad B_{\mu}^{(k)}(x,0)=\delta_{k 1}A_{\mu}(x)\, .
\label{eq:bexp}
\end{equation}
The flow field satisfies the same boundary conditions as the original gauge potential and can be Fourier expanded, at any given order, in the same way:
\be
B_\mu^{(k)} (x,t) = V^{-\half} \underset{q}{\sum^{\prime}} e^{i q x} \hat B_\mu^{(k)} (q,t)  \, \hat{\Gamma}(q)\, .
\label{eq:bexpfour}
\ee

The expansion of the energy density in powers of $g_0$ can now be obtained by expanding the fields in $E(t)$ directly. Dropping for clarity the arguments of the fields in position space, $B_{\mu}^{(n)}\equiv B_{\mu}^{(n)}(x,t)$, one gets, up to order $g_0^4$: 
\begin{align}
E(t)&= g_0^2 \, \Tr \left(\partial_\mu B_\nu^{(1)} \partial_\mu B_\nu^{(1)} - \partial_\mu B_\nu^{(1)} \partial_\nu B_\mu^{(1)}\right)
\label{eq:edens0}\\
&+ 2 i g_0^3 \, \Tr \left( \partial_\mu B_\nu^{(1)} \left[B_\mu^{(1)},B_\nu^{(1)}\right]\right)\nonumber \\
&+2 g_0^3 \, \Tr \left(\partial_\mu B_\nu^{(1)} \partial_\mu B_\nu^{(2)} -\partial_\nu B_\mu^{(1)} \partial_\mu B_\nu^{(2)}\right)\nonumber \\
&+ g_0^4 \, \Tr \left(\partial_\mu B_\nu^{(2)} \partial_\mu B_\nu^{(2)} - \partial_\mu B_\nu^{(2)} \partial_\nu B_\mu^{(2)} \right)\nonumber \\
&- \half g_0^4 \, \Tr \left( \left[B_\mu^{(1)},B_\nu^{(1)}\right]^2 \right)\nonumber \\
&+ 2 i g_0^4 \, \Tr \left(\partial_{\mu}B_{\nu}^{\left(2\right)}\left[B_\mu^{(1)},B_\nu^{(1)}\right] 
+ \partial_\mu B_\nu^{(1)}\left[B_\mu^{(1)},B_\nu^{(2)}\right] + \partial_\mu B_\nu^{(1)} \left[B_\mu^{(2)},B_\nu^{(1)}\right]\right)\nonumber \\
&+ 2 g_0^4 \, \Tr \left(\partial_\mu B_\nu^{(3)} \partial_\mu B_\nu^{(1)} - \partial_\mu B_\nu^{(3)} \partial_\nu B_\mu^{(1)} \right)+ \mathcal{O}(g_0 ^5). \nonumber
\end{align}

The corresponding expression in momentum space, however, is specific to the TGF set-up. In particular, the $SU(N)$ structure constants $f^{abc}$ appearing in infinite volume are replaced by 
the momentum dependent functions $F(p,q,r)$ appearing in the commutation relations of the $\hat{\Gamma}(q)$ matrices -- see eqs.~\eqref{eq:comm}, \eqref{eq:Fdef}.
For the sake of completeness we give below the seven different terms contributing to the expectation value of $\langle E(t)\rangle$ arising at order $g_0^4$, with an additional $1/N$ normalization factor added for later convenience. Each term can be identified with one of the lines in eq.~\eqref{eq:edens0}:
\begin{align}
\cE_0(t) &= \frac{g_0^2}{2NV} \sum_q^{\prime} \left(q^2\delta_{\mu\nu}-q_\mu q_\nu \right) \left\langle \B_\mu^{(1)}(-q,t)\,  \B_\nu^{(1)}(q,t) 
 \right\rangle 
\, ,
\label{eq:e0}\\
\cE_1(t) &= -\frac{g_0^3}{N V^{3/2}} \sum_{p_1,p_2,p_3}^{\prime} \!  \! \!   \delta\left(\sum p_i\right) F(p_1,p_2,p_3)\, i p_{1\mu}
 \label{eq:e1}\\
&\times\left\langle \B_\nu^{(1)}(p_1,t)\, \B_\mu^{(1)}(p_2,t)\, \B_\nu^{(1)}(p_3,t) \right\rangle 
\, ,
\nonumber
\\
\cE_2(t) &= \frac{g_0^3}{N V} \sum_q^{\prime} \left( q^2 \delta_{\mu\nu}-q_\mu q_\nu \right) \left\langle \B_\mu^{(1)}(-q,t)
\,\B_\nu^{(2)}(q,t) \right\rangle 
\, ,
\label{eq:e2}
\end{align}
\begin{align}
\cE_3(t) &= \frac{g_0^4}{2N V}\sum_q^{\prime} \left(q^{2}\delta_{\mu\nu}-q_{\mu}q_{\nu}\right)
\left\langle \B_\mu^{(2)}(-q,t)\, \B_\nu^{(2)}(q,t) \right\rangle 
\, ,
\label{eq:e3}\\
\cE_4(t) &= \frac{g_0^4}{4 N V^2}  \sum_{p_1,p_2,p_3,p_4}^{\prime}  \delta\left(\sum p_i\right) \,  F(p_1,p_2,-p_1-p_2)\,  F(p_3,p_4,-p_3-p_4) 
\label{eq:e4}\\
&\times \left\langle \B_\mu^{(1)}(p_1,t)\, \B_\nu^{(1)}(p_2,t)\, \B_\mu^{(1)}(p_3,t) \B_\nu^{(1)}(p_4,t) \right\rangle \, ,  \nonumber 
\\
%
\cE_5(t) &=
-\frac{ i g_0^4}{N V^{3/2}} \sum_{p_1, p_2, p_3}^{\prime} \! \! \! \!     \delta \left(\sum p_{i}\right)\,  p_{1\mu}\,  F(p_1,p_2,p_3)  
\Big \{
 \Big \langle   \B_\nu^{(2)} (p_1,t)\, \B_\mu^{(1)}(p_2,t)\, \B_\nu^{(1)} (p_3,t) \Big \rangle
\label{eq:e5}\\
&+ \Big \langle \B_{\nu}^{(1)}(p_{1},t)\, \B_{\mu}^{(2)}(p_{2},t)\, \B_{\nu}^{(1)}(p_{3},t) \Big \rangle + \Big \langle \B_{\nu}^{(1)}(p_{1},t)\, 
\B_{\mu}^{(1)}(p_{2},t)\, \B_{\nu}^{(2)}(p_{3},t)\Big \rangle  \Big \}  \, ,\nonumber  
\\
\cE_6(t) &=\frac{g_0^4}{N V}\sum_q^{\prime} \left(q^{2}\delta_{\mu\nu}-q_{\mu}q_{\nu}\right)\left\langle \B_{\mu}^{(1)}(-q,t)\, \B_{\nu}^{(3)}(q,t)\right\rangle 
\, .
\label{eq:e6}
\end{align}
The shorthand notation $\sum p_{i}$ in the $\delta$ functions was used to denote the sum over all present momenta for each term. The $\cE_{0}$ term will turn out to be a combination of a leading $\mathcal{O}\left(g_{0}^{2}\right)$ term and an $\mathcal{O}\left(g_{0}^{4}\right)$ correction, whereas all other terms will turn out to be $\mathcal{O}\left(g_{0}^{4}\right)$.

Then, the next step is to relate the flow fields to the actual gauge fields $A_{\mu}\left(x\right)$, for which we will need to obtain an order-by-order solution to the flow equations.

\subsubsection{Solving the flow equations in the TGF scheme}

Let us consider the flow equation~\eqref{eq:modflow} with the gauge parameter $\xi$ set to unity.  This was already solved by  L\"uscher for the infinite volume case \cite{Luscher:2010iy}, but the results in finite volume are slightly
different. Expanding the fields in perturbation theory, as in eq.~\eqref{eq:bexp}, and dropping for clarity of notation the arguments of the fields in position space, the equations to solve order by order are of the form:
\begin{equation}
\partial_{t}B_{\mu}^{\left(i\right)}=\partial_{\nu}^{2}B_{\mu}^{\left(i\right)}+R_{\mu}^{\left(i\right)},\qquad i\in \mathbb{Z} .
\end{equation}
The first three orders will be enough to obtain the observable at order $\mathcal{O}\left(g_{0}^{4}\right)$:
\begin{align}
R_{\mu}^{(1)}&=0 ,\\
R_{\mu}^{(2)}&=2i\left[B_{\nu}^{(1)},\partial_{\nu}B_{\mu}^{(1)}\right]-i\left[B_{\nu}^{(1)},\partial_{\mu}B_{\nu}^{(1)}\right],\\
R_{\mu}^{\left(3\right)}&=
-\left[B_{\nu}^{\left(1\right)},\left[B_{\nu}^{\left(1\right)},B_{\mu}^{\left(1\right)}\right]\right]+2i\left[B_{\nu}^{\left(1\right)},\partial_{\nu}B_{\mu}^{\left(2\right)}\right]-i\left[B_{\nu}^{\left(1\right)},\partial_{\mu}B_{\nu}^{\left(2\right)}\right]\\
&+ 2i\left[B_{\nu}^{\left(2\right)},\partial_{\nu}B_{\mu}^{\left(1\right)}\right]-i\left[B_{\nu}^{\left(2\right)},\partial_{\mu}B_{\nu}^{\left(1\right)}\right].
\nonumber
\end{align}
We may define a momentum space version of $R_{\mu}^{\left(i\right)}$:
\begin{equation}
R_{\mu}^{\left(i\right)}\left(x,s\right)= V^{-\half}\underset{p}{\sum^{\prime}}e^{ipx}R_{\mu}^{\left(i\right)}\left(p,s\right)\hat{\Gamma}\left(p\right),
\end{equation}
under which the $R_\mu$ terms read:
\begin{align}
R_\mu^{(1)}(p,t) &= 0, \\
R_\mu^{(2)}(p,t) &= \frac{i}{\sqrt V} \sum^{\prime}_q F(q,p,-q-p) \, \B_\nu^{(1)}(p-q,t) \Big (2 q_\nu \B_\mu^{(1)}(q,t) - q_\mu \B_\nu^{(1)}(q,t) \Big ),
\end{align}
\begin{align}
R_\mu^{(3)}(p,t) &=
 V^{-1}   \sum^{\prime}_{q_1,q_2,q_3} \delta\left(p-\sum_i q_i\right)  F(q_1,p,-q_1-p) F(q_2,q_3,-q_2-q_3) \, \\
&\times  \B_\rho^{(1)}(q_1,t) \B_\rho^{(1)}(q_2,t)  \B_\mu^{(1)}(q_3,t) \nonumber  \\
&- 2 i V^{-\frac{1}{2}}  \sum^{\prime}_{q_1,q_2}  \delta\left(p-\sum_i q_i\right) F(q_1,q_2,-q_1-q_2)\, \B_\rho^{(1)}(q_1,t) \B_\sigma^{(2)}(q_2,t) \nonumber \\
&\times \Big(q_{2\rho} \delta_{\sigma \mu} - 
q_{1\sigma} \delta_{\rho \mu} - \half (q_2-q_1)_\mu \delta_{\rho \sigma}\Big)  .
\nonumber
\end{align}
In terms of $\B_{\mu}\left(q,t\right)$, the flow equation in momentum space becomes:
\begin{equation}
\partial_{t}\B_{\mu}^{\left(i\right)}\left(p,t\right)= -p^{2}\B_{\mu}^{\left(i\right)}\left(p,t\right)+R_{\mu}^{\left(i\right)}\left(p,t\right),
\end{equation}
whose solution is immediate at first order:
\be
\B_{\mu}^{\left(1\right)}\left(p,t\right)=e^{-p^{2}t}\B_{\mu}^{\left(1\right)}\left(p,0\right)=e^{-p^{2}t}\A_{\mu}\left(p\right),
\ee
and which can be solved for the next two orders by directly integrating $R_{\mu}^{\left(i\right)}$:
\be
\B_{\mu}^{\left(i\right)}\left(p,t\right)=\int_{0}^{t}dse^{-\left(t-s\right)p^{2}}R_{\mu}^{\left(i\right)}\left(p,s\right), \quad i>1.
\ee
Higher order terms, while increasingly tedious, can be obtained through the same iterative procedure.

From these expressions, and using the Feynman rules from appendix \ref{ap:frules}, we derived the expressions of the contributions from eqs.~\eqref{eq:e0}-~\eqref{eq:e6} in terms of sums over momenta. Introducing for the sake of readability the symbol:
\be
 \wtsum \equiv \tl^{-2 d}\, \sum_{q,r}  NF^2(q,r,-q-r),
\label{eq:hatsum}
\ee
and after quite a bit of algebra, we ended up with:
\begin{align}
\cE_{0}(t) =   &\half \lambda_{0} \, \tl^{-d} \, \sum^{\prime}_q e^{-2tq^{2}}(d-1)
\label{eq:e00}
\\
+ & \frac{1}{4} \lambda_{0}^2 \, \wtsum e^{-2tq^2}  \frac{1}{q^2 r^2 (q+r)^2 } \left((3d-2) q^2 - 2 (d-2)^2 r^2\right),
\label{eq:e01}\\
\cE_{1}(t)\,  =   & \frac{3}{2} \lambda_{0}^2  \,\wtsum e^{-t(q^2+r^2+p^2)}\frac{1}{q^2 r^2} (1-d),
\label{eq:ee1}\\
\cE_{2}(t)\,  = &
 \lambda_{0}^2 \, \int_0^t ds \wtsum   e^{-2tp^2 + 2 s (qr)} \frac{1}{p^{2}q^{2}r^{2}}
\label{eq:ee2}\\
&\times \left\{(d-1) p^2 \left(p^2+q^2+r^2\right) + 2(d-2)\left(q^2 r^2-(qr)^2\right)\right\},
\nonumber
\end{align}
\begin{align}
\cE_{3}(t)\, = &
\lambda_{0}^2 \,\int_0^t ds s \wtsum  \, e^{-2t p^2}\left ( e^{2s(qr)}+e^{2(2t-s)(qr)}\right) \frac{1}{q^2 r^2}
\label{eq:ee3}\\
&\times \left\{2(d-1)p^2 r^2 + (d-2)\left(q^2 r^2-(qr)^2\right)\right\},
\nonumber\\
\cE_{4}(t)\, =& \frac{1}{4} \lambda_{0}^2\, \wtsum  e^{-2t (q^2+r^2)} \frac{1}{q^2 r^2} d(d-1),
\label{eq:ee4}\\
\cE_{5}(t)\, = & \lambda_{0}^2 \, \int_0^t ds \wtsum  e^{-(t+s)(q^2+r^2)-(t-s)p^2}
\frac{1}{q^2 r^2} (1-d) \left(5r^2+(qr)\right),
\label{eq:ee5}\\
\cE_{6}(t)\, = &
-\lambda^2_{0} \,  \int_0^t ds  \wtsum  e^{-2tq^2-2sr^2} \frac{1}{r^2} (d-1)^2 
\label{eq:ee6}\\
&+2 \lambda^2_{0} \, \int_0^t ds_1 \int_0^{s_1} ds_2  \wtsum  
 e^{-2tp^2+ 2 s_1 (qr)-2 s_2 (pq)} \frac{1}{p^2q^2} \nonumber\\ 
& \times \left\{2(d-2)\left(p^2q^2-(pq)^2\right)+(d-1)p^2 \left(2p^2-(qr)\right)\right\}, \nonumber
\nonumber
\end{align}
where the bare coupling and the volume have been replaced by the bare 't Hooft coupling and the effective length, and we have defined an auxiliary momentum $p=q+r$. The primes from the sums in the ${\cal O} (\lambda_0^2)$ terms have been discarded, as the $F^{2}(q,r,-q-r)$ factors automatically vanish for such momenta.

\subsubsection{The energy density at LO}
\label{s:LO}

As we recall, our aim in this paper was to obtain a perturbative expansion of the observable $\langle E(t)/N \rangle$ at NLO, which in powers of the bare 't Hooft coupling can be parametrized as:
\be
\left \langle \frac{E(t)}{N}\right \rangle \equiv\lambda_0 \, \cE^{(0)}(t)+\lambda_0^2 \, \cE^{(1)}(t)+\mathcal{O}\left(\lambda_0^3\right)\, .
\ee
We will begin by deriving the leading order term from the formulas in the previous subsection, given by:
\be
\cE^{(0)}(t)=   \half  (d-1)\, \tl^{-d} \, \sum^{\prime}_{m\in\mathbb{Z}^d} e^{- 8 t \pi^2   m^2/ \tl^2 }
\, .
\ee
It will be convenient to introduce a few auxiliary variables and functions:
\begin{align}
& t' \equiv 8 t /(c \tl)^2 \label{eq:tprime}, \qquad \hat{c} = \pi c^2/2,
\\
& {\cA}(x) \equiv  x^{d/2} \sum^{\prime}_{m\in\mathbb{Z}^d} e^{-\pi x  m^2 }\, , \label{eq:calA0}
\end{align}
in terms of which we may write:
\be
\cE^{(0)}(t) = 
  \frac{(d-1)}{2\,  (8\pi t) ^{d/2}}\, \, {\cA}(2\hc t').
\label{eq:order0}
\ee
This $\cA$ function can be expressed in terms of Jacobi theta functions $\theta_3$ as:
\begin{align}
{\cA}(x) &= x^{d/2} \ \theta_3^{(d-d_t)} ( 0 , i  x ) \left \{ \theta_3^{d_t}  ( 0 ,i  x) - \theta_3^{d_t} ( 0 ,i  x  l_g^2)  \right \}\, , \label{eq:calA} \\
{\cA}(x) &= \theta_3^{(d-d_t)} \left( 0 , \frac{i}{  x} \right) \left \{ \theta_3^{d_t}  \left( 0 ,\frac{i}{ x}\right) - \frac{1}{N^2} \theta_3^{d_t} \left( 0 ,\frac{i}{ x  l_g^2}\right)  \right \}\,, 
\label{eq:calAP}
\end{align}
where we used Poisson resummation to rewrite the theta functions:
\be
\theta_3 ( z ,i x)  = \sum_{m \in \mathbb{Z}} \exp \left \{ - \pi x m^2 + 2 \pi i m z \right \} = \frac{1}{\sqrt{x}} \sum_{m \in \mathbb{Z}} \exp \left \{ - \pi \frac{(m-z)^2}{x} \right \}  .
\label{eq:poisson}
\ee
The leading order infinite volume expression is retrieved in the $c \rightarrow 0$, $\tl \rightarrow \infty$ limit, taken in such a way as to keep $c \tl$ fixed. In that limit:
\be
\cA (2 \hc t') \rightarrow   \frac{N^2-1}{N^2} ,
\ee
leading to:
\be
\cE^{(0)}_\infty(t) =
  \frac{(d-1)(N^2-1)}{2\,  (8\pi t) ^{d/2}N^2}\, ,
\ee
in agreement with the results in ref.~\cite{Luscher:2010iy}.

\subsubsection{The energy density at NLO}
\label{s:integral}

As for the subleading ${\cal O}(\lambda_0^2)$ term coming from eqs.~\eqref{eq:e01} - \eqref{eq:ee6}, we found that, after a fair bit of algebra, it can be expressed in terms of a handful of integrals.
By rewriting the momenta in denominators as exponents using Schwinger's parametrization, and the momenta in numerators as derivatives with respect to the flow time variables, we were able to recast the expression for the energy density at NLO as:
\begin{align}
\cE^{(1)}(t)=& 2(d-2)\left(I_{1}+I_{2}\right)-4(d-1)I_{3}+4(3d-5)I_{4}+6(d-1)\left(I_{5}-I_{6}\right)
\label{eq:order1}
\\
-& 2(d-2)(d-1)I_{7}+ \half (d-2)^2  (  I_{8}+ 2 I_{9} ) -2(d-1) \left(I_{10}+I_{11}\right)-4(d-1)I_{12}
\, ,
\nonumber
\end{align}
where the $I_i$ are twelve relatively simple integrals to be detailed later on. As the computations and manipulations are rather long and tedious, we will illustrate the procedure using one of the simplest contributions, $\cE_4$ in eq.~\eqref{eq:ee4}, and show the remaining $\mathcal{E}_i$ contributions in terms of the basic integrals in appendix \ref{ap:intform}. The $\cE_4$ contribution is given by:
\be
\cE_{4}=\frac{1}{4}\lambda_{0}^{2} d\left(d-1\right){\wtsum} e^{-2t\left(q^{2}+r^{2}\right)}\frac{1}{q^{2}r^{2}}
\, .
\ee
Using Schwinger parametrization to lift the momenta from the denominator, we then defined an integral:
\be
I=\half{\wtsum}\int_{0}^{\infty}dz z \int_{0}^{1}dx \, e^{-\left(2t+x z\right)\, q^{2}-\left(2t+ (1-x) z\right)\, r^{2}}
\, ,
\ee
so as to rewrite:
\be
\cE_{4}=\half \, \lambda_{0}^{2} \, d(d-1)\, I
\, ,
\ee
with the structure constants entering this expression through the definition of the symbol $\wtsum$ given in eq.~\eqref{eq:hatsum}. 

The presence of the structure constants in each $\cE_{i}$ will let us formulate the integrands in terms of Siegel theta functions. We may indeed rewrite $N F^2$ as:
\be
NF^2 (q,r,-q-r)=1-\half \left(e^{i\theta_{\mu\nu} q_\mu r_\nu}+ e^{-i\theta_{\mu\nu}q_\mu r_\nu}\right)
\, ,
\ee
a substitution under which a generic integrand of the form:
\be
\cI={\wtsum}e^{-\ts q^{2}-\tu r^{2}-2 \tv q r}
\, ,
\ee
becomes:
\be
\cI=\tl^{-2d} \sum_{m,n\in\mathbb{Z}^d} {\rm Re}\left \{ e^{-\pi \left(s m^2+u n^2+ 2v m n\right)}
\left(1-e^{i 2 \pi \hat{\theta}n\tilde{\epsilon}m}\right)\right \} 
\, , 
\ee
where we rescaled the variables $s \equiv 4 \pi \tl^{-2} \ts$, $u \equiv 4 \pi \tl^{-2} \tu$, $ v  \equiv 4 \pi \tl^{-2} \tv$, and where we used the quantization of momenta in the twisted finite box to rewrite $q$ and $r$ in terms of integers.

The connection to Siegel theta functions becomes clear by introducing the function
\be
G(s,u,v,\htheta)\equiv \sum_{M\in\mathbb{Z}^{2d}}  { \rm Re } \left(e^{-\pi M^{t}A(s,u,v,0) M} -
e^{-\pi M^{t}A (s,u,v,\htheta) M}\right)\, , 
\ee
with 
\begin{align}
M=\left(\begin{array}{c}
m\\
n
\end{array}\right), \qquad A\left(s,u,v,\htheta\right)=\left(\begin{array}{lc}
s\mathbb{I}_{d} & v\mathbb{I}_{d}+i\hat{\theta}\tilde{\epsilon}\\
v\mathbb{I}_{d}-i\hat{\theta}\tilde{\epsilon} & u\mathbb{I}_{d}
\end{array}\right) \, .
\label{eq:adef}
\end{align}
In this expression $\mathbb{I}_{d}$ denotes the $d\times d$ identity matrix and
the sum over $M$ denotes the sum over the corresponding integers
$m_{\mu},n_{\nu}$, regrouped into a $2d$-dimensional column vector. 
Recalling the definition of the Siegel theta functions: 
\be
\Theta\left(z|A\right)\equiv\underset{M\in\mathbb{Z}^{2d}}{\sum}e^{i\pi\left(M^{t}AM+2z\cdot M\right)},
\ee
this matricial expression takes the form:
\be
G(s,u,v,\htheta)=\text{Re }\left\{\Theta(0|iA(s,u,v,0))-\Theta(0|iA(s,u,v,\htheta))\right\}
\, .
\label{eq:gdef}
\ee
Using this notation, the integral entering $\cE_4$ reads:
\be
I=\half\, \tl^{-2d}\int_{0}^{\infty}dz z \int_{0}^{1}dx \,  G\left(4 \pi \tl^{-2} \left(2t+x z\right),4\pi\tl^{-2} \left(2t+(1-x) z\right),0,\htheta\right)
\, .
\ee

With this, only one last bit of manipulation is left in order to have the integrals ready for the calculation of the energy density at NLO. In terms of the variables $t'$ and $\hc$ defined in eq.~\eqref{eq:tprime} and rescaling $z$ appropriately we have:
\be
I=\frac{\hat c^2}{32 \pi^2 \tl^{2d-4}}\int_{0}^{\infty}dz z \int_{0}^{1}dx \,  G\left(\hat c \left(2t'+x z\right),\hat c \left(2t'+(1-x) z\right),0,\htheta\right)
\, .
\ee 
Introducing an auxiliary function $\Phi(s,u,v,\htheta)$ that incorporates the normalization factor in front of the integral:
\be
\Phi(s,u,v,\htheta)= {\cal N} G(\hat c s, \hat c u,\hat c v,\htheta), \label{eq:phi}
\ee
with:
\be
\cN = \frac{\hc^2}{32 \pi^2  \tilde{l}^{2d-4} }\, , \label{eq:calN}
\ee
we rewrote the integrals in a fairly basic form allowing us to evaluate them numerically. For instance, for the integral in $\cE_4$: 
\begin{equation}
I=\int_{0}^{\infty}zdz\int_{0}^{1}dx\, \Phi(2t'+xz,2t'+\left(1-x\right)z,0,\htheta)\, .
\end{equation}

A similar procedure can be followed for all the terms contributing to the energy density at NLO, leading to the result in eq.~\eqref{eq:order1} for $\cE^{(1)}(t)$, where the twelve intervening integrals are:
\begin{align}
I_{1}(\Phi,t')= & \int_{0}^{t'}dxx \, \Phi(2t',2x,x,\htheta)\, ,
\label{eq:i1}\\
I_{2}(\Phi,t')= & \int_{0}^{t'}dx x\, \Phi(2t',2t',x,\htheta)\, ,
\label{eq:i2}\\
I_{3}(\Phi,t')= & \int_{0}^{\infty}dz t' \Phi(2t'+z ,2t',t',\htheta) ,
\label{eq:i3}\\
I_{4}(\Phi,t')= & \int_{0}^{t'}dx x\int_{0}^{1}dy\, \Phi(2t',2x,xy,\htheta)\, ,
\label{eq:i4}\\
I_{5}(\Phi,t')= & \int_{0}^{\infty}dz\int_{0}^{t'}dx x\, \Phi(2t',(z+2)x,x,\htheta)\, ,
\label{eq:i5}\\
I_{6}(\Phi,t')= & \int_{0}^{\infty}dz\int_{0}^{t'}dx \, \Phi(2t'+z,2t',x,\htheta)\, ,
\label{eq:i6}
\\
I_{7}(\Phi,t')= & \int_{0}^{\infty}dz\int_{0}^{t'}dxx\, \Phi(2t',(z+2)x,0,\htheta)\, ,
\label{eq:i7}
\\
I_{8}(\Phi,t')= & \int_{0}^{\infty}dz\int_{0}^{\infty }dy\, \Phi(2t'+z,2t'+y,0,\htheta)\, ,
\label{eq:i8}
\\
I_{9}(\Phi,t')= & \int_{0}^{\infty}z dz\int_{0}^{\infty}dy\int_{0}^{1}dx\, \partial_{z'}\Phi(2t'+x z+y,z',xz,\htheta)\Big|_{z'=z}\, ,
\label{eq:i9}
\\
I_{10}(\Phi,t')= & \int_{0}^{\infty}dz\int_{0}^{t'}dx x\, \partial_{t'}\Phi(2t'+z,2t',x,\htheta)\, ,
\label{eq:i10}
\\
I_{11}(\Phi,t')= & \int_{0}^{\infty}dz\int_{0}^{t'}dx x^2\, \partial_{t'}\Phi(2t',(z+2)x,x,\htheta)\, ,
\label{eq:i11}
\\
I_{12}(\Phi,t')= & \int_{0}^{\infty}dz\int_{0}^{1}dy\int_{0}^{t'}dxx^2\, \partial_{t'}\Phi(2t',(z+2)x,x y,\htheta)\, .
\label{eq:i12}
\\ \nonumber
\end{align}

\subsection{Structure of UV divergences}
\label{s:diver}

As some of the integrals defined in the previous subsection are UV divergent in 4 dimensions, in this subsection we will discuss how to parametrize their asymptotic behavior. We will show that, in all of our cases, the divergent contributions can be expressed in terms of an infinite volume integral that can be regularized through analytic continuation in $d$. The relation to the existing infinite volume calculation from ref.~\cite{Luscher:2010iy} will be presented in section~\ref{s:largev}.

The UV singularities are tied to the structure of the Siegel theta functions entering the definition of the $\Phi$ function: 
\begin{equation}
\Theta\left(0|iA(\hc s,\hc u,\hc v,\htheta)\right)=\underset{m,n\in\mathbb{Z}^d}{\sum}\exp\left(-\pi\hat{c}\left(sm^{2}+un^{2}+ 2 v mn\right)-2\pi i\hat{\theta}m\tilde{\epsilon}n\right)\, .
\label{eq:siegeldiv}
\end{equation}
The real part of the matrix $A(\hc s,\hc u,\hc v,\htheta)$, obtained by setting $\htheta=0$ in eq.~\eqref{eq:adef}, is a positive definite symmetric matrix as long as $\det A(\hc s, \hc u, \hc v,\htheta= 0)>0$, i.e. when $(su-v^2)>0$, which ensures that the series defining the theta function converges uniformly. It will be useful to define a new quantity:
\be
\alpha = s-\frac{v^2}{u}\, ,
\label{eq:alpha}
\ee
which is always positive definite in our integration ranges, and hence the determinant will be positive definite as well, except at the points for which $u=0$.~\footnote{In some cases divergences happen for the points $(s,u,v)=(2t',2t',2t')$, but, using a particular momentum shift, they can be moved to $(s,u,v)= (2t',0,0)$. Indeed, the substitution: $ u^{\prime}=s+u-2v,\, v^{\prime}=v-s$ can be implemented by shifting $m\rightarrow m-n$ within the momentum sums. Such a shift had already been applied in order to give the definitions of the integrals in the previous subsection.}

The analysis of the asymptotic behavior of the integrals is much clearer once we apply Poisson resummation as in eq.~\eqref{eq:poisson} to each component of $n$ in the definition of $\Theta$:
\begin{equation}
\Theta\left(0|iA\left(\hc s,\hc u,\hc v,\htheta\right)\right)=\left(\hat{c}u\right)^{-\frac{d}{2}}\underset{m,n\in\mathbb{Z}^d}{\sum}\exp\left(-\pi\hat{c}sm^{2}
-\frac{\pi}{\hat{c}u}\left(n-\hat{\theta}\tilde{\epsilon}m-i\hat{c}vm\right)^{2}\right)
\, .
\end{equation}

Whenever $\htheta  \tilde \epsilon m \notin \mathbb{Z}^d$, the corresponding term will be asymptotically finite at $u=0$. However, in the case in which we have a vector of integers, we will be able to remove the $\htheta$ dependence by shifting $n$, thus leaving the asymptotic behavior to be driven by the shifted $n=0$ terms. Such terms go, as we approach $u\rightarrow0$, as:
\be
\left(\hat{c}u\right)^{-\frac{d}{2}}\underset{m\in\mathbb{Z}^d}{\sum}\exp\left(-\pi\hat{c} \alpha m^2\right)\, .
\ee
This observation allows us to isolate the asymptotic divergence by identifying the cases for which $\htheta  \tilde \epsilon m \in \mathbb{Z}^d$. The first case in which this occurs is whenever $\htheta\equiv\bar k / l_g =0$, for any value of $m$. For nonzero $\htheta$, it will happen whenever $\tilde \epsilon m = 0 $ (mod $l_g)$. Since the vector $\tilde \epsilon m$ has nonvanishing components only along the twisted directions, this will be the case whenever  $m_\mu = 0 $ (mod $l_g)$ simultaneously for all twisted directions.

The terms responsible for the UV divergences at $u=0$ have therefore been identified, and come in two categories:
\begin{itemize}
\item For $\hat{\theta}=0$, terms with $n=0$ and any value of $m$. 
\item For $\hat{\theta}\neq0$, terms with $n=0$ (after shifting away the $\htheta$ dependence), and $m_\mu=0\text{ (mod }l_{g})$ in all twisted directions at once.
\end{itemize}
With this, we may begin the discussion on how the divergent integrals can be regularized.

\subsubsection{Regularization}
\label{s:regularization}

The regularization strategy will be based on splitting each integral into the sum of a finite piece that can be directly evaluated at $d=4$ and integrated numerically, and an asymptotic term to be handled analytically using dimensional regularization. The way to implement such a strategy will be discussed in this subsection.

We start by introducing a function $H(s,u,v,\htheta)$ given by:
\begin{equation}
H(s,u,v,\htheta)={\cal N} \,  \underset{n\in\mathbb{Z}^d}{\sum} \underset{m\in\mathbb{Z}^d}{\sum}^{\prime}\text{Re } \left \{\exp\left(-\pi\hat{c}\left(sm^{2}+un^{2}+2v mn\right)
-2\pi i\hat{\theta}m\tilde{\epsilon}n\right) \right \}
\, ,
\label{eq:Hdef}
\end{equation}
with the usual meaning for the prime in the sum over $m$. The $\Phi$ function entering the integrals can then be rewritten as:
\begin{equation}
\Phi(s,u,v,\htheta)= H(s,u,v,0)-H(s,u,v,\htheta)
\, ,
\label{eq:Phidef}
\end{equation}
which is quite advantageous, as the explicit exclusion of the momenta $m$ proportional to $l_g$ from the sum automatically makes the term in $\htheta \ne 0$ finite at $u=0$. All UV divergences at $d=4$ thus come, in this parametrization, from the $H(s,u,v,0)$ term, and are of the form:
\be
\Phi^{(0)} (s,u,v) ={\cal N} \left(\hat{c}u\right)^{-\frac{d}{2}}\underset{m\in\mathbb{Z}^d}{\sum}^{\prime}\exp\left(-\pi\hc \alpha m^{2}\right)
\, ,
\label{eq:tphi0}
\ee
with $\alpha$ defined as in eq.~\eqref{eq:alpha}, and, as we recall, positive definite everywhere in the integrals. Hence, the sum over $m$ is convergent, and the leading asymptotic behavior at $u=0$ is controlled by the $u^{-d /2}$ factor (times the additional powers of $u$ appearing in the integrand prefactor). It will be useful to write the function $\Phi^{(0)}$ in terms of the function $\cA(x)$ from eq.~\eqref{eq:calA0}:
\be
\Phi^{(0)}(s,u,v) =
{\cal N}  \hc^{-d} \left (u \alpha\right)^{-d /2} \, \cA\left(\hc \alpha\right) \, .
\ee
For reasons that will become clear later, we define:
\be
\Phi^\infty(s,u,v) = {\cal N}  \hc^{-d}  \left (u \alpha\right)^{-d /2},
\label{eq:phinfty}
\ee
in terms of which we may rewrite:
\be
\Phi^{(0)}(s,u,v) = \cA\left(\hc \alpha\right)  \Phi^\infty(s,u,v) \, .
\label{eq:phi02}
\ee

This formulation will be useful to analyze the asymptotic UV behavior of the integrals resulting from replacing the original function $\Phi$ in the integrand by $\Phi^{(0)}$. Before discussing the general treatment, we will deal with $I_{1}$ as a representative example. In this case the integral diverges at $x=0$ as $u=2x$. The piece containing the divergence thus reads:
\be
I_1 (\Phio,t') = 
\int_{0}^{t'}dx x \,  \Phio \left(2t',2x,x\right) = \int_{0}^{t'}dx x \,  \Phi^\infty \left(2t',2x,x\right) {\cA} \left(2\hc t'- \hc x/2\right) ,
\ee
which, substituting in the expression for $\Phi^\infty$, yields:
\be
I_1 (\Phio,t') ={\cal N}  \hc^{-d}  \int_{0}^{t'}dx \, x^{1-d/2} (4t'-x)^{-d /2} {\cA} \left(2\hc t'- \hc x/2\right) \, .
\ee
The asymptotic behavior at small $x$ can then be obtained by expanding ${\cA} \left(2\hc t'- \hc x/2\right)$ around $x=0$. The integrand of the leading term goes as $x^{1-d/2}$, whereas the next to leading term is convergent in $d=4$. Hence, the integral will behave asymptotically as:
\be
I_1^\rd (t') =     {\cA}(2\hc t')   \int_{0}^{t'}dx x \, \Phi^\infty\left (2t',2x,x\right)\equiv {\cA}(2\hc t') \,  I_1(\Phi^\infty, t')\, . 
\ee
Notice that in this expression the entire momentum dependence has been factorized into the normalization constant ${\cA}(2\hc t')$, which happens to be the same factor that appeared at leading order -- see eq.~\eqref{eq:order0}. The integral $ I_1(\Phi^\infty,t')$ can then be evaluated in dimensional regularization with $d=4-2\epsilon$, leading to:
\be
I_1(\Phi^\infty,t') =  {\cal N} \hc^{-d}
 \int_{0}^{t'}dxx^{1-d/2}
(4t'-x)^{-d/2} =  \frac{{\cal N}\hc^{-d} }{4} \, (2t')^{2-d} \left ( \frac {1}{\epsilon} +  \frac {1}{3} + \log \frac {4}{3} \right )
\, .
\ee
The asymptotic expansion of all other integrals (except for $I_9$, which we will address separately) is obtained in the same way: we expand the function $\cA(\hc \alpha)$ appearing in the definition of $\Phi^{(0)}$ around $u=0$, retain the leading term, and then use it to define:
\be
I_i^\rd (t') = {\cA}(2\hc t') \, I_{i } (\Phi^\infty,t') .
\label{eq:phi0div}
\ee
Remarkably, the integrals $I_i(\Phi^\infty,t')$ match the ones appearing in the infinite volume calculation (up to a factor depending on $N$), which we will present in sec.~\ref{s:largev}.

We are now in a position to summarize, still keeping $I_9$ aside, the regularization strategy. The idea is to decompose the finite volume integrals into two pieces, one that is finite in four dimensions:
\be
I_i^\rf (t') =  I_i(\Phi-\Phio, t') + I_i(\Phio, t')  - \cA(2\hc t') \, I_i (\Phi^\infty,t') \, ,
\label{eq:Ifin}
\ee
and another one, shown in eq.~\eqref{eq:phi0div} above, that requires analytic continuation to four dimensions and is proportional to each corresponding infinite volume integral. The ultraviolet divergences of the original integral are contained in this last piece, and appear as poles in $1/(d-4)$, though only $I_1,I_4,I_5$ and $I_7$ turned out to have such $1/\epsilon$ poles.

As for the strategy to regularize $I_9$, some modifications, described in detail in appendix~\ref{ap:I9}, are required. The initial integral is decomposed as:
\be
I_9 (t') =  I_9(\Phi-\theta(1-z) \Phio, t')  - I_9(\theta(z-1) \Phio, t') +  I_9(\Phio, t') \, ,
\ee
with the Heaviside step function $\theta$ restricting the interval of integration over $z$. The first term on the right hand side is finite in four dimensions, while the other two have to be analytically continued to $d=4$. Denoting $I_9^\text{reg}$ these analytic continuations, we end up with: 
\begin{align}
&I_9^{\rm reg}(\Phio, t')=0 \, , \\
&I_9^{\rm reg}(\theta(z-1)\Phio, t') = -{\cal N} \hc^{-4} \int_{0}^{\infty}dz\, \left \{
  (2t'+z)^{-2}  \cA\left(\hc(2t'+z)\right) \right . \label{eq:deltaI9}\\
&\left .+ 
 \int_{0}^{1}dx\, \left(2t'+x(1-x)+z\right)^{-2} \cA\left(\hc(2t'+x(1-x)+z)\right)
  \right \}
\, .
\nonumber
\end{align}
And therefore:
\be
I_9^\rf (t') =  I_9(\Phi-\theta(1-z) \Phio, t')  -  I_9^{\rm reg} (\theta(z-1) \Phio, t')\, ,
\label{eq:I9fin}
\ee
and:
\be
I_9^\rd (t') = 0 \, .
\ee

\subsection{Infinite volume limit}
\label{s:largev}

The expression of the energy density in infinite volume can be easily retrieved (see \cite{Perez:2017jyq}) by making the following substitutions in eqs. \eqref{eq:e00} - \eqref{eq:ee6}:
\begin{align}
&\tl^{-d} \sum_q'  \longrightarrow   {\frac{N^2-1}{N^2}}   \int \frac{d^dq}{(2\pi)^d} \, ,
\label{eq:infvol1}\\
&\wtsum\equiv \tl^{-2d}  \sum_{q,r}  N F^2(q,r) \longrightarrow {\frac{N^2-1}{N^2}} \int \frac{d^dq}{(2\pi)^d}\int \frac{d^dr}{(2\pi)^d}
\,.
\label{eq:infvol2}
\end{align}
The resulting expressions for the contributions to the energy density, after integrating over the d-dimensional momenta, can once again be rewritten in terms of twelve basic integrals, much like what happened in the finite volume case. 
We will first present the case of $\cE_4$ as an illustrative example, and then present the results for the general case.

The  infinite volume expression for $\cE_4$  is obtained making the substitutions from eqs.~\eqref{eq:infvol1} and ~\eqref{eq:infvol2} in eq.~\eqref{eq:ee4}. After integrating over momenta, we have:
\begin{equation}
\cE_{4}^\infty(t) =\frac{\lambda_{0}^{2}\, d\left(d-1\right) (N^2-1) }{4 N^2(4\pi)^d} \left(\int_{0}^{\infty}dz (2t+z)^{-d/2} \right)^2
\, .
\end{equation}
Setting $t= \hc \tl^2 t' /(4\pi)$ and recalling the definition of $\Phi^\infty$ from~\eqref{eq:phinfty}, one trivially derives:
\begin{equation}
\cE_{4}^\infty(t) = \frac{\lambda_{0}^{2} }{2} \, d\left(d-1\right) \frac{N^2-1 }{ N^2} \, I_8(\Phi^\infty,t') \, .
\end{equation}
Comparing this with the finite volume expression:
\begin{equation}
\cE_{4}=\frac{\lambda_{0}^{2}}{2} \, d\left(d-1\right)  I_8(\Phi,t') \equiv \frac{\lambda_{0}^{2}}{2} \, d\left(d-1\right)  \left (I_8^\rf(t') + \cA(2\hc t') \, I_8 (\Phi^\infty,t') \right) \, ,
\end{equation}
we can relate the finite and infinite volume expressions for $\cE_{4}$. In fact, the infinite volume expression can be obtained from the finite volume one by taking the $\hat{c}\rightarrow 0$, $\tilde{l} \rightarrow \infty$ limit
at fixed $t'$, as $I_8^\text{fin}(t')$ vanishes and $\mathcal{A}(2\hc t')$ becomes $\frac{N^2-1}{N^2}$. A detailed discussion on that limit can be found in section~\ref{s:colours}.

Similar results hold for the other integrals, and thus the infinite volume energy density can be reproduced by performing a simple change in the finite volume formula from eq.~\eqref{eq:order1}:
\begin{equation}
I_i(\Phi, t') \longrightarrow \frac{N^2-1 }{ N^2}\,  I_i^\infty(t') \, ,
\end{equation}
where $I_9^\infty(t') = 0$ (see appendix~\ref{ap:I9}), and $I_i^\infty(t') = I_i (\Phi^\infty,t')$ for the rest.

Computing the infinite volume integrals in dimensional regularization with $d=4-2\epsilon$, one derives the energy density:
\begin{equation}
\left \langle \frac{E(t)}{N} \right \rangle = \frac{\lambda_0(N^2-1) (d-1) }{ 2 N^2  (8\pi t)^{d/2}} \left \{ 1 + \lambda_0 \frac{(8t)^\epsilon (4 \pi)^\epsilon}{ 16 \pi^2 } \left ( \frac{11}{3\epsilon}
+  \frac{52}{9} -3 \log 3 \right) \right \} \, ,
\label{eq:luscher_res}
\end{equation}
which agrees with the result obtained by L\"uscher in ref.~\cite{Luscher:2010iy}.

\section{'t Hooft coupling at one-loop}
\label{s:oneloop}

As we provided in the previous section a regularized expression for the expectation value of the energy density, we are now finally able to focus on several interesting results. Namely, we will in this section derive the running of the coupling, its relation to the $\MS$ coupling, obtain the $\Lambda$ parameter, and present our numerical results for the case of the $d_t=2$ two-dimensional twist.

\subsection{Perturbative matching to the \texorpdfstring{$\MS$}{MS} coupling at one-loop order}

Let us begin by recapitulating what has been achieved so far. As we recall, we expanded the observable $\langle E(t)/N \rangle$ up to NLO in powers of the 't Hooft coupling:
\be
\left \langle \frac{E(t)}{N}\right \rangle \equiv\lambda_0 \, \cE^{(0)}(t)+\lambda_0^2 \, \cE^{(1)}(t)+\mathcal{O}\left(\lambda_0^3\right)\, ,
\ee
with the leading order term being given by:
\be
\cE^{(0)}(t) =  \frac{(d-1) {\cA}(2 \hc t')}{ 2 (8\pi t)^{d/2}} \, .
\ee  
The function $\cA(x)$ was defined in eq.~\eqref{eq:calA}, and the variables $\hc =\pi c^2/2$ and $t'=8t/(c\tl)^2$ were introduced to make the expression more compact.

The NLO contribution is written in terms of twelve integrals given in eqs.~\eqref{eq:i1}-~\eqref{eq:i12}, regulated through analytic continuation in $d=4-2\epsilon$. The leading asymptotic behavior of each integral was identified, and a subtraction procedure was implemented, allowing us to write the energy density at NLO as:
\be
\cE^{(1)}(t)= \cE_\rf^{(1)}(t)+ \cE_\rd^{(1)}(t)\, .
\ee
All of the $1/\epsilon$ poles arising in dimensional regularization are contained in $\cE_\rd^{(1)}(t)$, a quantity that can be trivially rewritten in terms of the infinite volume result $\cE_\infty^{(1)}(t)$ as:
\be
\cE_\rd^{(1)}(t) = \frac{N^2 {\cA}(2\hc t')}{N^2-1} \cE_\infty^{(1)}(t) \, .
\ee
  
Gathering all of these pieces, our results for the expectation value of the energy density can be summarized in the following expression:
\be
\left \langle \frac{E(t)}{N} \right \rangle = \lambda_0  \, \cE^{(0)}(t)
\left \{ 1 + \lambda_0 \frac{(8t)^\epsilon (4 \pi)^\epsilon}{ 16 \pi^2 } \left ( \frac{11}{3\epsilon}
+\frac{52}{9} -3 \log 3 +  C_1 (t) \right ) + \mathcal{O}(\lambda_0^2) \right \} ,
\ee
where  $C_1(t) $ is given by:
\be
C_1(t)=   
\frac{16 \pi^2}{\cE^{(0)}(t)} \,   \cE^{(1)}_\rf(t) \, .
\ee
The perturbative relation to the $\MS$ coupling at one-loop order is obtained by simply introducing the expression of the bare coupling in terms of the $\MS$ one:
\be
\lambda_0 = \lambda_\MS \, \mu^{2\epsilon}  \left (4 \pi e^{-\gamma_E}\right)^{-\epsilon} 
 \left \{ 1 -  \frac{\lambda_\MS}{ 16 \pi^2 } \, \frac{11}{3\epsilon} \right \} \, ,  
\ee
leading to:
\be
\left \langle \frac{E(t)}{N} \right \rangle = \cE^{(0)}(t) \lambda_\MS \left \{1 + \frac{\lambda_\MS}{ 16 \pi^2 }
\left ( \frac{11}{3} \log(8t \mu^2 e^{\gamma_E})  + \frac{52}{9} -3 \log 3 +
C_1 (t) \right)  \right \} .
\ee
Setting the $\MS$ scale to $\mu = 1/\sqrt{8t} = 1/(c \tl)$, the relation at one-loop order between the two couplings reads:
\be
\lambda_\TGF(\tl) = \lambda_\MS(\mu)  \left \{1 + c_1 \lambda_\MS(\mu)  \right \} ,
\ee
with the following matching coefficient at one-loop order:
\be
c_1= \frac{1}{16 \pi^2} \left(\frac{11}{3} \gamma_E +\frac{52}{9} -3 \log 3 + \cC_1 \right)\, ,
\ee
and where we introduced the one-loop constant $\cC_1$:
\be
\cC_1 = C_1\left(t=c^2 \tl^2/ 8\right) .
\ee
The ratio between $\Lambda$ parameters in both schemes is then determined, as usual, in terms of the finite one-loop constant $c_1$:
\be
\log \left (\frac{\Lambda_\TGF}{\Lambda_\MS} \right) = \frac{3}{22} \left (\frac{11}{3} \gamma_E + \frac{52}{9} -3 \log 3 +
\cC_1\right) = \frac{c_1}{2b_0}  \, .
\ee 
The purpose of the rest of this section will be to evaluate $\cC_1$ numerically, in the case of a single non-trivially twisted plane.

\subsection{The matching coefficient for a two-dimensional twist}
\label{s:results}

The ingredients required in order to compute the finite constant $\cC_1$, entering the ratio $\Lambda_\TGF/\Lambda_\MS$, have been provided in sec.~\ref{s:diver}. In the specific case of $d_t=2$, the computational effort that has to be invested in order to determine $\cC_1$ is considerably smaller than for $d_t=4$, as the $8 \times 8$ matrices entering the expression for $\Phi$ are reduced to, at most, $4\times 4$. In particular, we have:
\begin{align}
H(s,u,v,\htheta) &= {\cal N} \Theta^{d-2}\left (0|iA_0\left(\hc s,\hc u,\hc v\right)\right ) \! \\ \nonumber
 & \times \text{Re} \left \{ \Theta\left (0|iB\left(\hc s,\hc u,\hc v,\htheta\right)\right )   -
 \Theta^2\left (0|iA_0 \left(\hc s l_g^2 ,\hc u,\hc v l_g \right)\right ) \right\}, \\
 \Phi^{(0)}(s,u,v) &= {\cal N} (\hc u)^{-d/2} \ \theta_3^{d-2} ( 0 , i \hc \alpha ) \left \{ \theta_3^{2}  ( 0 ,i \hc \alpha )
- \theta_3^{2} ( 0 ,i \hc \alpha  l_g^2)  \right \}\, ,
\end{align}
where we defined a $2\times 2$ matrix $A_0$:
\begin{align}
A_0\left( s, u, v\right)=\left(\begin{array}{lc}
 s & v\\
 v &  u
\end{array}\right) \, ,
\label{eq:a0def}
\end{align}
as well as a $4\times 4$ matrix $B$ containing the $\htheta$ dependence given by, denoting $\epsilon$ the two-dimensional Levi-Civita symbol:
\begin{align}
B\left( s, u, v,\htheta\right)=\left(\begin{array}{lc}
A_0 ( s, u, v)& -i\hat{\theta}\epsilon\\
i\hat{\theta}\epsilon & A_0  ( s, u, v)
\end{array}\right) \, .
\label{eq:bdef}
\end{align}

The starting point for the numerical calculation of $\cC_1$ will then be given by eqs.~\eqref{eq:Ifin} and ~\eqref{eq:I9fin}, defining $I_i^\rf$. All these integrals have been built to be finite, so $d$ can be set to four, and
$l_g$ to $N$, in all intervening expressions. The calculation will come in two steps, the first of which will involve using a short Mathematica program to evaluate:
\begin{align}
&I_i(\Phio, t'=1)  - \cA(2\hc) \, I_i (\Phi^\infty,t'=1) & \text{for }  i=1,\cdots 8 \text{ and  }  i=10, \cdots 12 ,\\
&I_9^{\rm reg} (\theta(z-1) \Phio, t'=1) .
\end{align}
The required Jacobi theta functions are part of the standard Mathematica package, and for the integration we used the numerical integrators provided by the program by default. The derivatives appearing in some of the integrals were computed using finite differences.

\begin{table}[t]
  \begin{center}
    \begin{tabular}{c|c||c|c||c|c||c|c}
$c$ & $\cC_1$ & $c$  & $\cC_1$  & $c$ & $\cC_1$  & $c$ & $\cC_1$   \\
\hline
0.18    & 0.224(8)  & 0.34   & 0.301(5)   & 0.50    & -1.831(3) &0.66  & -4.577(12)\\
0.19    & 0.289(7)  & 0.35   & 0.228(6)   & 0.51    & -2.014(6) &0.67  & -4.727(20)\\
0.20    & 0.353(7)  & 0.36   & 0.142(2)   & 0.52    & -2.198(3) &0.68  & -4.862(17)\\
0.21    & 0.404(6)  & 0.37   & 0.0530(16) & 0.53    & -2.383(3) &0.69  & -4.998(17)\\
0.22    & 0.451(2)  & 0.38   &-0.0464(5)  & 0.54    & -2.569(3) &0.70  & -5.119(15)\\
0.23    & 0.493(2)  & 0.39   &-0.152(3)   & 0.55    & -2.755(4) &0.71  & -5.239(10)\\
0.24    & 0.536(3)  & 0.40   &-0.268(2)   & 0.56    & -2.947(8) &0.72  & -5.359(10)\\
0.25    & 0.557(3)  & 0.41   &-0.385(2)   & 0.57    & -3.125(9) &0.73  & -5.460(17)\\
0.26    & 0.570(2)  & 0.42   &-0.525(4)   & 0.58    & -3.303(10)&0.74  & -5.581(14)\\
0.27    & 0.567(4)  & 0.43   &-0.664(3)   & 0.59    & -3.482(10)&0.75  & -5.705(10)\\
0.28    & 0.558(3)  & 0.44   &-0.813(2)   & 0.60    & -3.646(14)&0.76  & -5.806(18)\\
0.29    & 0.532(5)  & 0.45   &-0.971(5)   & 0.61    & -3.808(10)&0.77  & -5.877(60)\\
0.30    & 0.508(4)  & 0.46   & -1.134(5)  & 0.62    & -3.968(12)&0.78  & -6.011(40)\\
0.31    & 0.473(4)  & 0.47   & -1.302(4)  & 0.63    & -4.125(16)&0.79  & -6.141(40)\\
0.32    & 0.426(6)  & 0.48   & -1.474(3)  & 0.64    & -4.280(12)&0.80  & -6.248(40)\\
0.33    & 0.361(3)  & 0.49   & -1.650(3)  & 0.65    & -4.435(16)&      &          \\
    \end{tabular}
  \caption{Results for $\cC_1$ for the $SU(3)$ gauge group and a range of values of $c$. }
    \label{t:SU3}
  \end{center}
\end{table}

\begin{figure}[ht]
\includegraphics[width=0.85\linewidth]{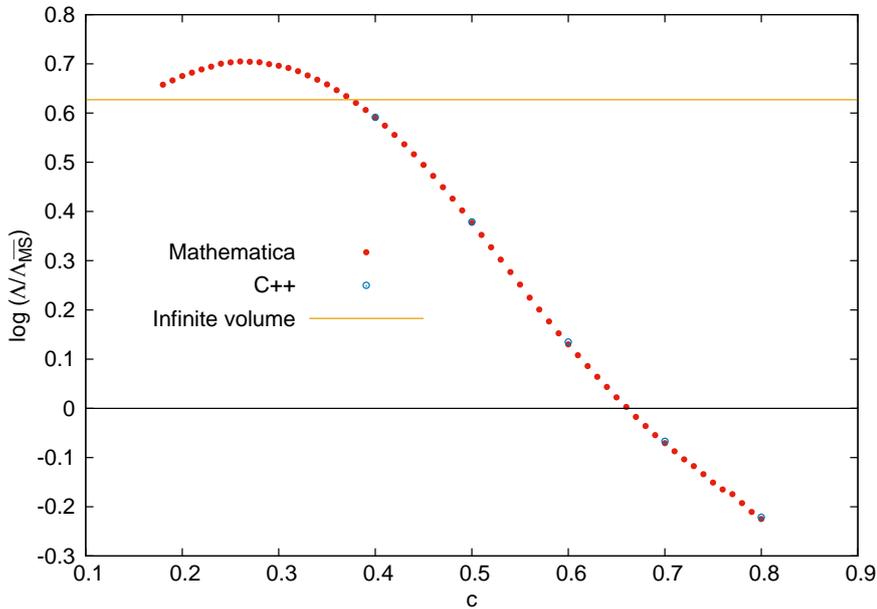}
\caption{ We display $\log(\Lambda / \Lambda_{\MS})$  as a function of $c$ for the $SU(3)$ gauge group and $\kb =1$.}
\label{f:SU3}
\end{figure}

The second step is far more complex from a numerical viewpoint, as it encompasses the calculation of:
\begin{align}
&I_i(\Phi-\Phio, t'=1) & \text{for }  i=1,\cdots 8 \text{ and  }  i=10, \cdots 12, \\
&I_9(\Phi-\theta(1-z) \Phio, t'=1) .
\end{align}
Two independent codes were prepared for this second step, one of them written in Mathematica~\footnote{In this case it turned out to be convenient to evaluate the two steps of the calculation jointly.} and the other in C++. The former, much like in the first step, made use of the standard Mathematica packages, numerical integrators, and finite differences to compute the integrals, whereas the full details of the inner workings of the latter can be found in appendix~\ref{ap:numerical}. 
We will simply mention here that different errors were used for each of the integrals, depending on computation time. The relative errors ranged from $10^{-8}$ in the best cases (for the single integrals), to $10^{-3}$ at worst for $I_9$, which was by far the bottleneck. The value of $c$ also had significant effects, with lower values taking longer times to compute.

Two key aspects are particularly interesting in the analysis of the results: the dependence on $c$ of the coupling at constant $\htheta$, and the general dependence in $\htheta$. 

For an example of the former, we analyzed in detail the case of $SU(3)$ with $\kb =1$, with $c$ ranging from 0.18 to 0.8. The results for $\cC_1$ are shown in table~\ref{t:SU3}. Figure~\ref{f:SU3} displays $\log(\Lambda_{\TGF}/\Lambda_{\MS})$ as a function of $c$. In a few points we plot the results obtained with both the Mathematica and the C++ codes, which are perfectly compatible (errors in the data points are smaller than the size of the symbol). The yellow horizontal line shows the result obtained when the gradient flow coupling is evaluated at infinite volume. A detailed analysis on the approach to the infinite volume and the dependence on the number of colors is presented in sec.~\ref{s:colours}, but for now we will simply mention that at constant energy scale $\mu= (c\tilde{l})^{-1}$ and fixed $N$, taking $c \rightarrow 0$ is equivalent to taking the large volume limit, in which $\log(\Lambda_{\TGF}/\Lambda_{\MS})$ should approach the yellow line in the plot. 

\begin{table}[t]
  \begin{center}
    \begin{tabular}{c|c|c||c|c|c|c|c|c}
      $\bar{k}$ & $N$ & $\hat{\theta}=\bar{k}/N$ & c=0.4 & c=0.5 & c=0.6 & c=0.7 & c=0.8  \\
      \hline
      1 & 7 & 0.1429 & -4.672(15) & -5.814(20) & -6.813(26) & -7.799(39) & -8.693(62) \\
      1 & 6 & 0.1667 & -4.274(14) & -5.729(19) & -6.979(24) & -8.097(35) & -9.080(47) \\
      1 & 5 & 0.2000 & -3.417(12) & -5.098(17) & -6.573(23) & -7.811(33) & -8.843(43) \\
      1 & 4 & 0.2500 & -2.049(12) & -3.808(16) & -5.475(22) & -6.833(30) & -7.912(40) \\
      2 & 7 & 0.2857 & -1.187(13) & -2.891(15) & -4.634(20) & -6.050(29) & -7.156(39) \\
      1 & 3 & 0.3333 & -0.261(14) & -1.818(14) & -3.614(19) & -5.087(29) & -6.220(38) \\
      3 & 8 & 0.3750 & 0.327(14) & -1.073(16) & -2.888(19) & -4.395(26) & -5.545(37) \\
      2 & 5 & 0.4000 & 0.583(14) & -0.724(16) & -2.542(19) & -4.064(26) & -5.222(37) \\
      3 & 7 & 0.4286 & 0.791(12) & -0.418(16) & -2.236(19) & -3.771(26) & -4.937(36) \\
     5 & 11 & 0.4545 & 0.911(09) & -0.228(15) & -2.045(19) & -3.587(26) & -4.757(36) \\
      1 & 2 & 0.5000 & 1.077(13) & -0.092(16) & -1.914(19) & -3.461(26) & -4.634(36) \\
    \end{tabular}
  \caption{Results for $\cC_1$ for several $SU(N)$ gauge groups and values of $\kb$.}
    \label{t:C1}
  \end{center}
\end{table}

\begin{figure}[ht]
\begin{center}
  \includegraphics[angle=-90,width=.90\linewidth]{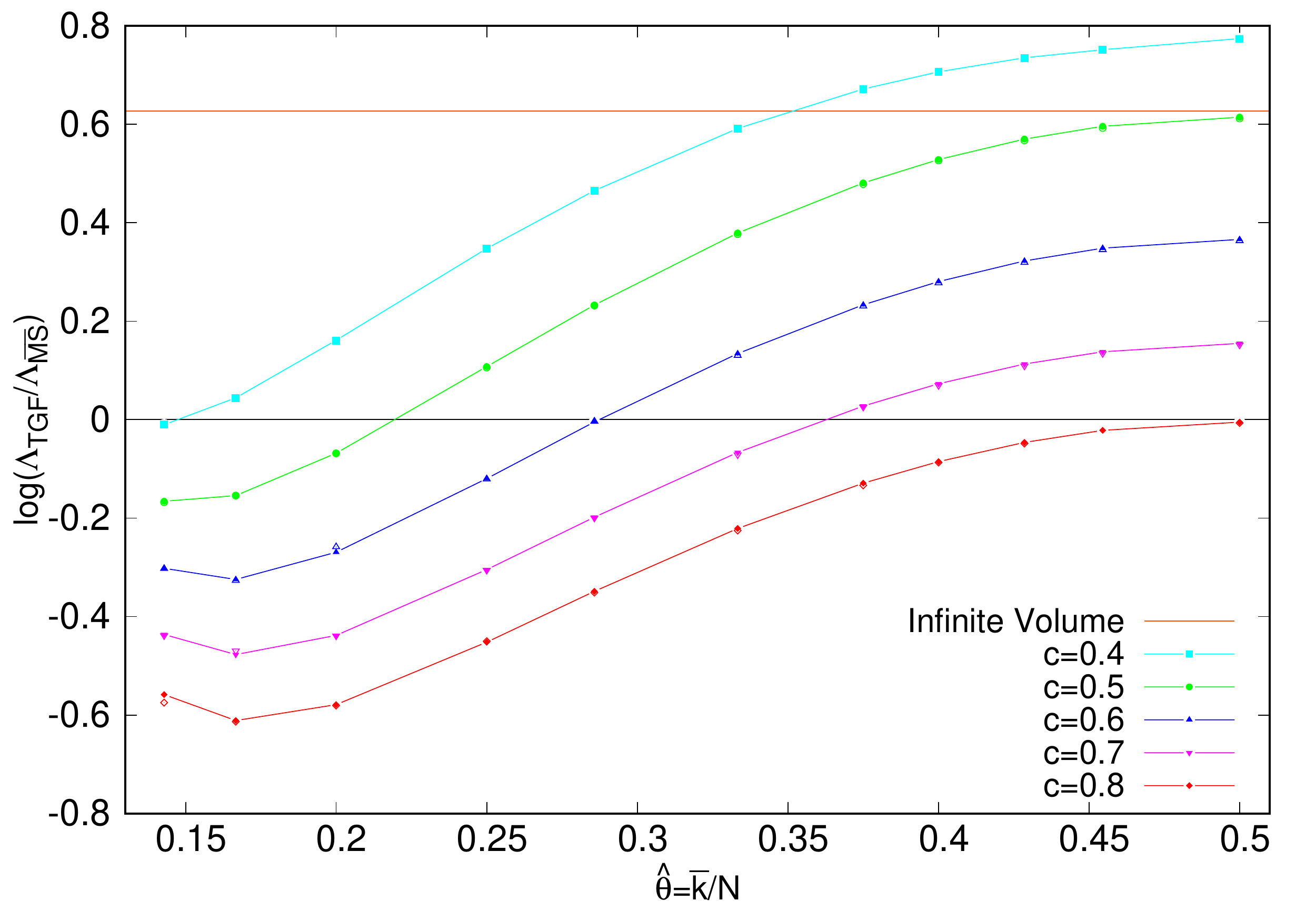}
\end{center}
  \caption{We display $\log(\Lambda_\TGF / \Lambda_{\MS})$ as a function of $\htheta= \bar k/N$ for different choices of gauge group and magnetic flux. The hollow and full symbols correspond to the results of the Mathematica and C++ codes respectively, though the overlap makes them indistinguishable in almost every case. The error bars of the results are smaller than the symbols used, but the errors can easily be worked out from table~\ref{t:C1}. }
  \label{f:Full}
\end{figure}

As for the study of the general dependence on $\htheta$, we considered a series of coprime values of $\bar{k}$ and (small) $N$ such that $\hat{\theta}$ ranged from 0.14 to 0.5. The full results for $\cC_1$ are shown in table~\ref{t:C1} and fig.~\ref{f:Full}, in which they are plotted as a function of $\htheta$ for several values of $c$. We observe that the dependence on $\htheta$ is rather smooth for the considered values of $\bar{k},N$. A discussion about the $\htheta$-dependence for larger values of $N$ will be presented in sec.~\ref{s:colours}.

\section{Dependence on the number of colors and the magnetic flux}
\label{s:colours}

In this section, we will analyze the dependence of $\lambda(c\tl)$ on the number of colors $N$ and the angular variable $\htheta= \bar k /l_g$. We will consider two different limits, both of them taken at fixed value of the renormalized 't Hooft coupling. 
The first is a singular large $N$ limit in the spirit of those introduced in ref.~\cite{AlvarezGaume:2001tv}, in which $N$ is sent to infinity while the torus size is sent to zero in such a way as to keep $\tl$ fixed, and the second is the thermodynamic limit, achieved by simultaneously sending $c$ to zero and $\tl$ to infinity while keeping $c \tl$ fixed. The idea that the infinite volume limit can be attained at $\tl \rightarrow \infty$ by sending either the torus size or the number of colors to infinity is implicit in our construction.

\subsection{Singular large \texorpdfstring{$N$}{N} limit and \texorpdfstring{$\htheta$}{theta}-dependence}

Singular large $N$ limits such as the one described above have been employed in various contexts. In ref.~\cite{GarciaPerez2015} the non-perturbative running of the $SU(\infty)$ 't Hooft coupling was computed through a step scaling procedure implemented by changing the rank of the gauge group. The calculation was done in the extreme case of TEK reduction on a one-site lattice with an effective size given by $\tl = a \sqrt{N}$, where $a$ denotes the lattice spacing. The continuum limit at fixed $\tl$ was achieved by sending $N$ to infinity, allowing the authors to compute the evolution of the coupling constant through a wide range of scales, and matching the two-loop perturbative formula at small coupling rather well.

These type of limits  have also been considered in the framework of non-commutative field theory. The gauge theory we are considering is equivalent, through the Morita duality, to a non-comutative gauge theory whose rational adimensional non-commutativity parameter is given precisely by $\htheta$, a mapping through which the effective torus size $\tl$ corresponds directly to the size of the non-commutative torus in the dual theory.
One of the proposals raised in ref.~\cite{AlvarezGaume:2001tv} was to define non-commutative gauge theories at irrational values of $\htheta$ through a sequence of ordinary $SU(N_i)$ twisted Yang-Mills theories with increasing number of colors and $\htheta_i = \kb_i/N_i \rightarrow \htheta$. In 2+1 dimensions, ref.~\cite{Perez:2018afi} has shown that this is only possible, avoiding tachyonic instabilities, for an uncountable  zero-measure set of values of $\htheta$, such as for instance a sequence of values of $\kb$ and $N$ defined through $\kb_i/N_i  = F_{i-2}/F_i$, where $F_i$ denotes the $i$th term in the Fibonacci sequence. In that case, instabilities in the large $N$ limit are avoided and the limiting sequence tends to $\htheta = (3-\sqrt{5})/2$.

In 2+1 dimensions, the condition required to avoid instabilities has been shown to be given in terms of a quantity dubbed $Z_{\min}$:
\be
Z_{\min} (N, k) = \min_{ m\ne 0 \, ({\rm mod}\,  l_g)} \, m \, ||\htheta m|| \, ,
\ee
where the symbol $||x||$ is used to denote the distance from $x$ to the nearest integer~\cite{Chamizo:2016msz,Perez:2018afi}. Tachyonic instabilities and symmetry breaking transitions can be avoided as long as $Z_{\min}>0.1$. Remarkably, this parameter also controls, in 4-dimensional perturbation theory, the size of the contribution of non-planar diagrams to the expectation value of Wilson loops~\cite{Perez:2017jyq}.

The limiting procedure to define non-commutative gauge theories at irrational values of the non-commutativity parameter relies on the asumption of continuity in $\htheta$. The one-loop matching constant $\cC_1$ depends on the choice of the parameter $c$ defining the renormalization scheme, the rank of the group, and the magnetic flux $k$, and, in particular, given a fixed value of $c$, one should analyze under which conditions the $k$ and $N$ dependence is fully encoded in the dimensionless ratio $\bar k/N$ defining $\htheta$. 
While a detailed analysis of the $\htheta$ dependence is beyond the scope of this paper, we did look at the integrals $I_1$ and $I_2$ entering the definition of $\cC_1$ as representative examples of integrals that are respectively UV divergent and finite after dimensional regularization. 

\begin{figure}[t]
\centering
\begin{subfigure}{.5\textwidth}
\raggedleft
  \includegraphics[width=1.\linewidth]{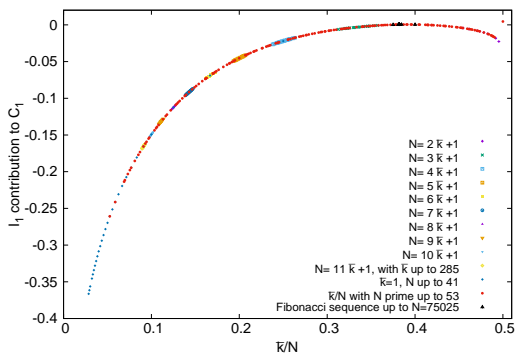}
\captionsetup{width=.9\textwidth}
  \caption{$ I_1$}
  \label{f:i1a_theta}
\end{subfigure}%
\begin{subfigure}{.5\textwidth}
\raggedleft
  \includegraphics[width=1.\linewidth]{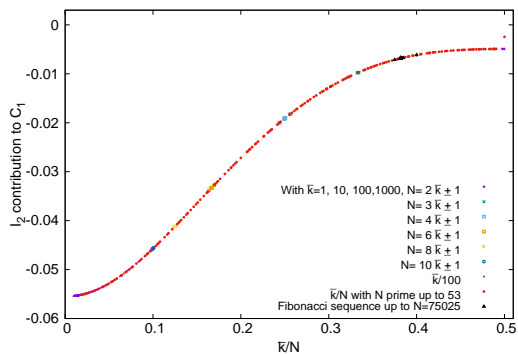}
\captionsetup{width=.9\textwidth}
  \caption{$ I_2$}
  \label{f:i2a_theta}
\end{subfigure}
\caption{Dependence on $\htheta$ of the $I_1$ and $I_2$ contributions to $\cC_1$ at $c=0.30$.}
\label{f:c15_theta}
\end{figure}
\begin{figure}[t]
\centering
\begin{subfigure}{.5\textwidth}
\raggedleft
  \includegraphics[width=1.\linewidth]{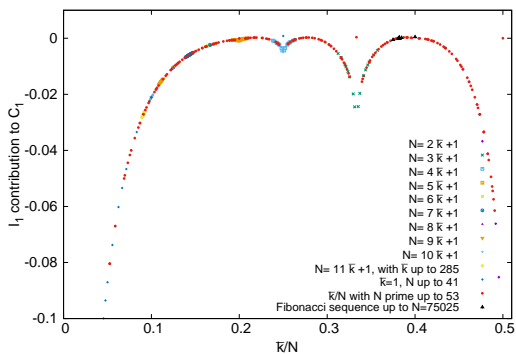}
\captionsetup{width=.9\textwidth}
  \caption{$ I_1$}
  \label{f:i1b_theta}
\end{subfigure}%
\begin{subfigure}{.5\textwidth}
\raggedleft
  \includegraphics[width=1.\linewidth]{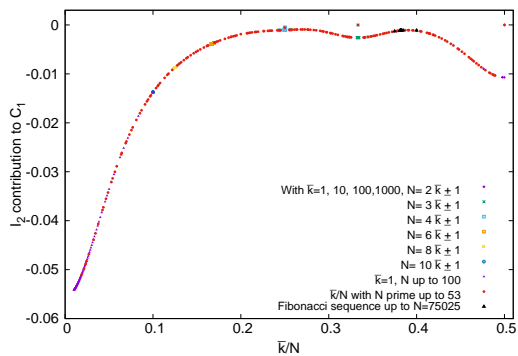}
\captionsetup{width=.9\textwidth}
  \caption{$ I_2$}
  \label{f:i2b_theta}
\end{subfigure}
\caption{Dependence on $\htheta$ of the $I_1$ and $I_2$ contributions to  $\cC_1$
at $c=0.15$.}\label{f:c30_theta}
\end{figure}

Figures~\ref{f:c15_theta} and ~\ref{f:c30_theta} show how the $I_1$ and $I_2$ contributions to $\cC_1$ depend on $\htheta$ for $c=0.15$ and $c=0.30$ respectively. We have explored many values of $N$ ranging from $N=2$ to $N=75025$, the latter as part of the aforementioned Fibonacci sequence. For $c=0.3$, we noticed that the dependence on $\htheta$ of both integrals is continuous, with the exception of the point $N=2$ in the case of $I_1$. As $c$ decreases, however, several other points corresponding to small values of $N$ deviate from the general curve, and, in the case of $I_1$, we observe a steep dependence on $\htheta$ for sequences approaching rational values, in particular for $\bar k /N= 0$, 1/4, 1/3 and 1/2. A similar dependence on $\htheta$ has been observed in lattice perturbation theory when considering the contribution at second order of non-planar diagrams to the expectation values of Wilson loops~\cite{Perez:2017jyq}, which can be understood in terms of the parameter $Z_{\rm min}$ introduced earlier.

\begin{figure}[t]
\centering
\begin{subfigure}{.5\textwidth}
\raggedleft
  \includegraphics[width=1.\linewidth]{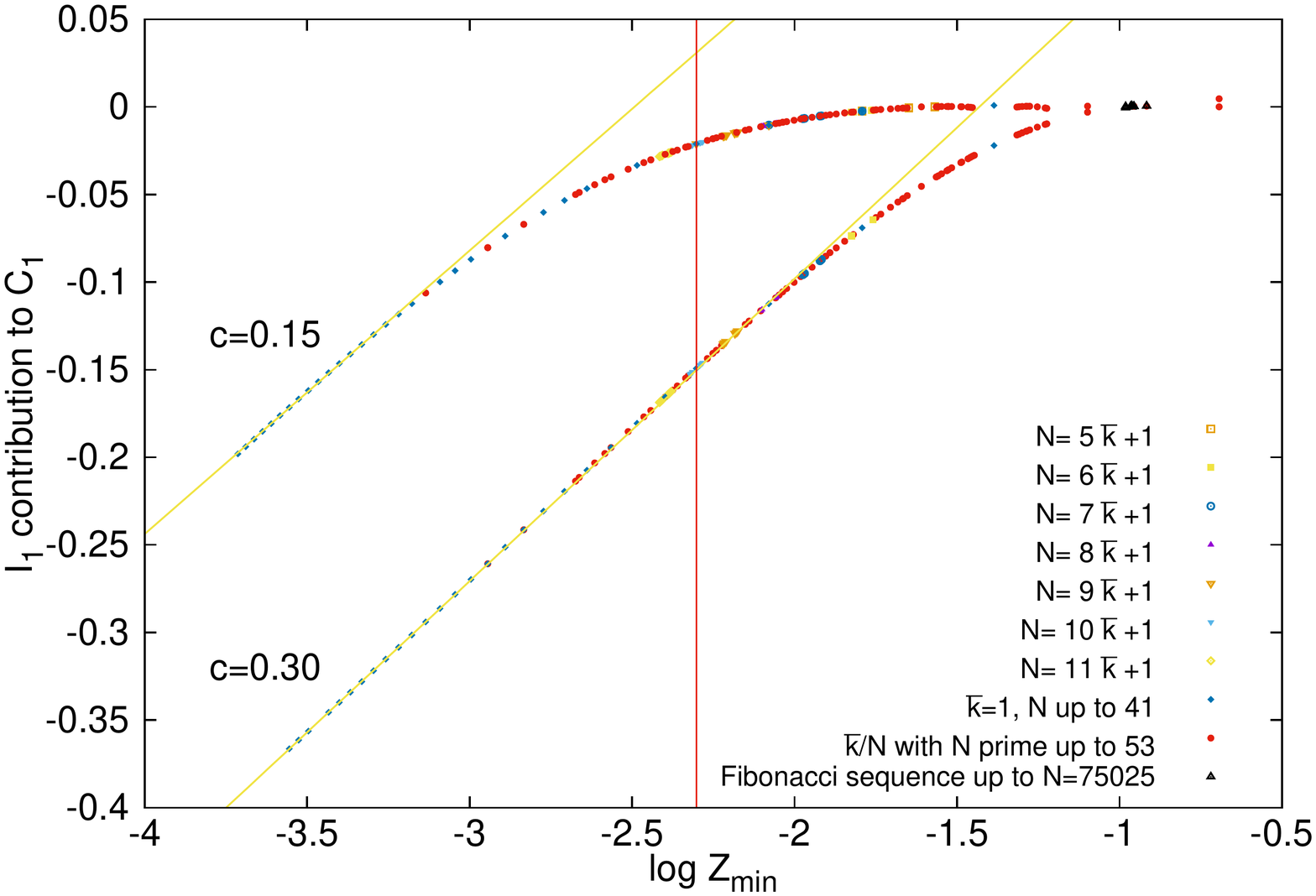}
\captionsetup{width=.9\textwidth}
  \caption{$ m_0=1$}
  \label{f:i1a_zmin}
\end{subfigure}%
\begin{subfigure}{.5\textwidth}
\raggedleft
  \includegraphics[width=1.\linewidth]{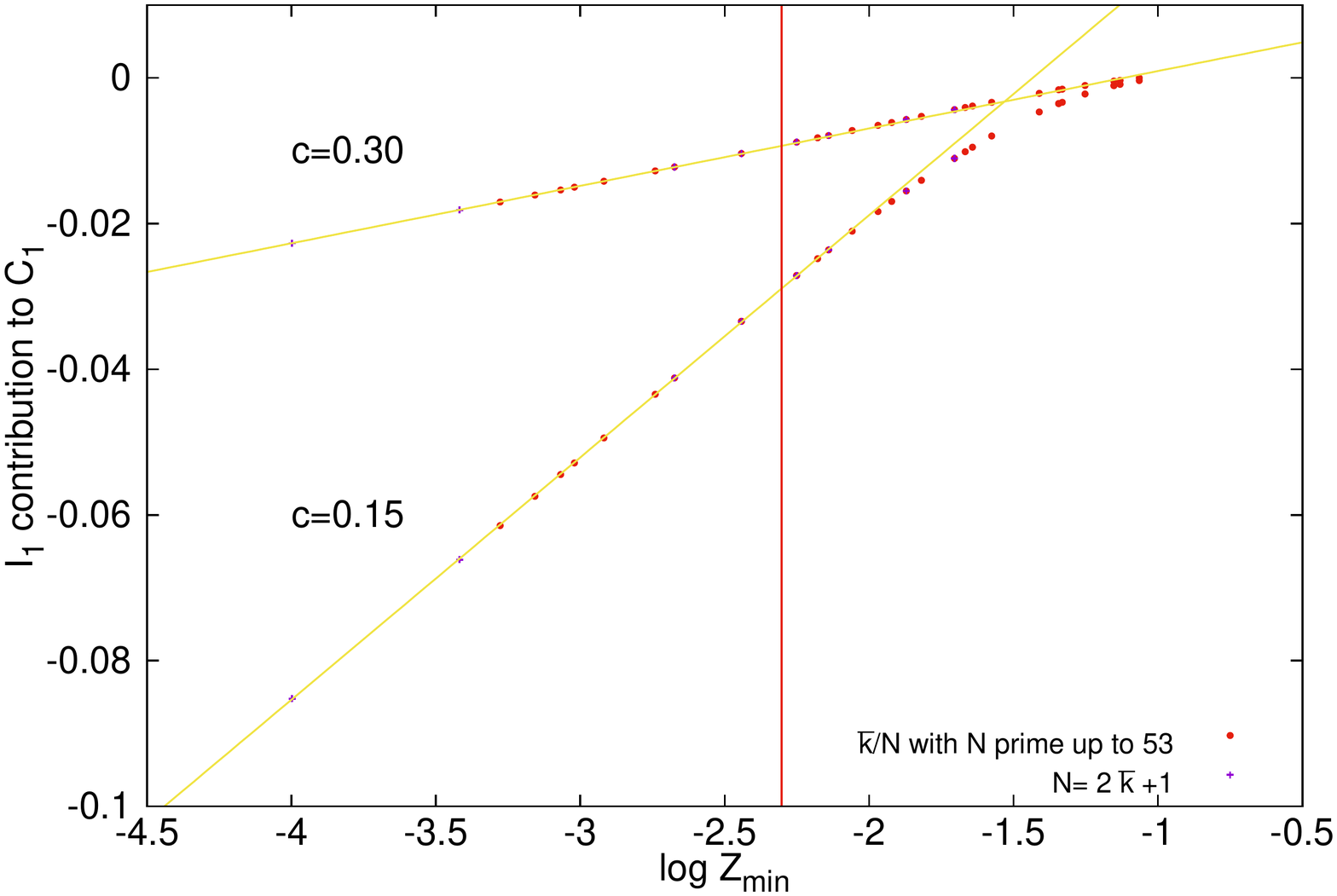}
\captionsetup{width=.9\textwidth}
  \caption{$ m_0=2$}
  \label{f:i1b_zmin}
\end{subfigure}
\caption{$Z_{\rm min}$-dependence of the contribution of $I_1$ to the one-loop matching constant $\cC_1$. The red vertical line in the plots corresponds to $Z_{\rm min}=0.1$.}
\label{f:i1_zmin}
\end{figure}

Let us take a look at how the dependence in this $Z_{\rm min}$ quantity enters in the $I_1$ contribution to $\cC_1$. The $\htheta$-dependent term comes from the function $H(s,u,v,\htheta)$ defined in eq.~\eqref{eq:Hdef}. This contribution is finite in the UV and given by:
\be
-\frac{\hc^2}{3 \cA (2 \hc)} \int_0^1 \frac{dx}{ x}  
 {\sum_{m\in \Z^4}}^\prime \sum_{n\in \Z^4}   \exp \left \{- \frac{\pi  \hc}{2} (4-x)   m^2 - {\frac{\pi}{2 \hc x}}
(n-\htheta  \tilde \epsilon m)^2  +  i \pi  \, m n \right\}
\, .
\ee
As all terms included in the sum have a non-zero value of $\htheta  \tilde \epsilon m$, UV-finiteness is guaranteed. However, in the limit in which this quantity tends to zero, one would retrieve the divergence present in the $\htheta=0$ term. We will in what follows show that such a limit is approached logarithmically in $Z_{\rm min}$. Let us begin by considering the leading asymptotic behavior for small $x$:
\be
-\frac{\hc^2 }{3 \cA (2 \hc)}  \theta_3^2\left (0,2i \hc \right) \sum_{m\in \Z^2}^\prime    e^{- 2 \pi  \hc m^2 + i  \pi  \, m \hat n } 
\int_0^1 \frac{dx}{ x}      \exp \left \{- {\frac{\pi}{2 \hc x}} ||\htheta \tilde{\epsilon}  m||^2  \right\}
\, ,
\ee
where $\hat n$ denotes the integer closest to $\htheta  \tilde \epsilon m$. Integrating over $x$, we get:
\be
-\frac{\hc^2 }{3 \cA (2 \hc)} \theta_3^2\left (0,2i \hc \right) \sum_{m\in \Z^2}^\prime    e^{- 2 \pi  \hc m^2 + i  \pi  \, m \hat n } 
\, \Gamma\left[0,\frac{\pi Z^2(m)}{2 \hc m^2}\right] ,
\ee
where $Z^2(m) =  m^2 \, ||\htheta  m||^2$.
If the argument of the incomplete $\Gamma$ function is small, this goes as:
\be
\frac{\hc^2}{3 \cA (2 \hc)} \theta_3^2\left (0,2 i \hc \right) \sum_{m\in \Z^2}^\prime    e^{- 2 \pi  \hc m^2 + i  \pi  \, m \hat n } 
\left (\gamma_E + \log \left (\frac{\pi Z^2(m)}{2 \hc m^2} \right) \right) 
+ \cdots
\label{eq:zdep}
\ee
The logarithmic dependence in $Z$ is tamed by the exponential damping in $\hc m^2$, but at small enough $\hc$ this suppression disappears, giving rise to the behavior presented in fig.~\ref{f:i1b_theta}. This is more clearly seen in fig.~\ref{f:i1_zmin} where we show the contribution of $I_1$ to $\cC_1$ as a function of $\log Z_{\rm min}(N,k)$. The left plot shows the points for which the minimal value is attained at $m= (1,0)$, and the right one those with the minimum at $m=(2,0)$, with the red vertical line in the plots corresponding to $Z_{\rm min}=0.1$. Sequences approaching $\htheta=0$ in the left plot and $\htheta=1/2$ in the right one are deep in the region with small $Z_{\rm min}$, where a tiny change in the value of $\htheta$ translates into a large change in the integral. 

As a final remark, we will point out that the value of $Z_{\rm min}$ stays almost constant along the Fibonacci sequence mentioned earlier, meaning that the results of the integrals will depend almost exclusively on the value of $c$. Therefore, as expected, the singular large $N$ limit can be taken safely along such a sequence, making it optimal, for instance, for the determination of the $SU(\infty)$ running coupling using the reduction techniques employed in ref.~\cite{GarciaPerez2015}. 

\subsection{Large volume limit}

So far, we have been discussing the dependence of the matching constant $\cC_1$ on the number of colors and the flux-dependent parameter $\htheta$ for a fixed value of $c$, the parameter defining the TGF scheme. In contrast, in this subsection we will be looking at a different type of limit, namely the one in which $c$ tends to zero while the effective size is sent to infinity in such a way as to keep flow time fixed (thus fixing the energy scale of the coupling as well). This limit can be taken in two different ways, either by sending the smallest torus period $l$ to infinity while keeping the rank of the group $N$ fixed, or by sending $N$ to infinity at fixed $l$. If volume independence holds true, in both cases the infinite volume expression should be recovered, and correspondingly $\cC_1$ should vanish. As we recall, at fixed value of $t$, $\cC_1$ is a function of three parameters: $c$, $N$ and the magnetic flux $k$. In particular, all of the dependence on the boundary conditions (i.e. the dependence on $k$) is contained in $\cC_1$, and will vanish in the thermodynamic limit provided $\cC_1$ does as well. We will therefore analyze in what follows the behavior of the matching constant in the approach to the thermodynamical limit, along with the size of the finite volume (or finite $N$) corrections.
 
To prepare for such a discussion, we will first take a look at the LO term in the expansion of the energy density, eq.~\eqref{eq:order0}, with $t$ set to $(c\tl)^2/8$. As we recall, the dependence on $c$ and $N$ came from:
\be
\cA (\pi c^2)  = F_0(\pi c^2,4-d_t) \left(F_0(\pi c^2,d_t)- \frac{1}{N^2} F_0(\pi c^2 l_g^2,d_t)\right)  
\, ,
\ee
where:
\be
F_0(x,d) = \sum_{m\in \mathbb{Z}^d} \exp\left(- \frac{\pi m^2 }{ x} \right) 
\, .
\ee
In the infinite volume limit, understood in the sense of $c \rightarrow 0$ at fixed $l_g$, one has $F_0(0,d)= 1$ and therefore:
\be
\cA (\pi c^2 )  \rightarrow  1  - \frac{1}{N^2}     
\, ,
\ee
leading to a LO term in agreement with the results found in ref.~\cite{Luscher:2009eq}. The leading correction is exponentially suppressed with the square of the volume as:
\be
-\frac{2 d_t}{N^2} \exp\{-1/(c l_g)^2\} \equiv  -\frac{2 d_t}{N^2} \exp\{-l^2/(8t)\} .
\ee 
If the large $N$ limit (i.e. large $l_g$) at fixed $l$ and constant $c l_g$ is taken instead, one gets $\cA (\pi c^2 )  = 1 + {\cal O} (1/N^2) $, which does indeed correspond to the infinite volume large $N$ limit. The approach to the limit is in that case powerlike, with $1/N^2$ corrections.

The discussion of the NLO term, on the other hand, is more involved and requires some previous steps to be properly considered. As we recall, the different contributions to $\cC_1$ can be written in a compact way as\footnote{The regularized expression for $I_9$ is slightly different, see app.~\ref{ap:I9}.}:
\be
\tI= \frac{4}{3 \cA (2 \hc)} \int (u \alpha)^{-2} \left(\hat H(s,u,v,0) - \hat H(s,u,v,\htheta) - \cA(2\hc)\right)
\, ,
\ee
where we used the symbol $\int$ to denote the integrals appearing in eqs.~\eqref{eq:i1}-\eqref{eq:i12} in a generic manner, including the prefactors multiplying the $\Phi$ function and derivatives when required. The quantity $\hat H(s,u,v,\htheta)$ is related to the function $H(s,u,v,\htheta)$ entering the definition of $\Phi$ through:
\be
H(s,u,c,\htheta) = \Phi^\infty(s,u,v) \hat H(s,u,v,\htheta)\, ,
\ee
and it is given by:
\be
 \hat H(s,u,v,\htheta) = {\rm Re} \left \{F_1(\alpha, u, v,0,4-d_t )\left (  F_1(\alpha, u, v,\htheta,d_t ) - \frac{1}{N^2} F_1(\alpha l_g^2, u, v l_g ,0,d_t) \right)
\right \} \, ,
\label{eq:hhat}
\ee
with:
\be
F_1( \alpha, u,v,\htheta,d) =  (\hc \alpha)^{d/2} \sum_{m,n\in \Z^d}   \exp \left \{- \pi  \hc \alpha  m^2 - {\frac{\pi}{\hc u}}
(n-\htheta  \tilde \epsilon m)^2  + 2 \pi i \frac{v}{u}  \, m n   \right\}
\, .
\ee

In order to analyze the approach to the infinite volume limit, it is more convenient to look at the expression resulting after Poisson resummation in $m$. We will, for simplicity's sake, focus on the case of the two-dimensional twist, $d_t=2$, and will move the full detail of the computations to appendix~\ref{ap:largeN} for clarity. We will separate each of the contributions to $\cC_1$ into $\htheta-$independent and $\htheta-$dependent terms, given by:
\begin{align}
I_{TI} &= I_{TI}^{(0)} + \frac{4}{3 \cA (2 \hc)} \left \{ \int (u \alpha)^{-2}  \hat H'(s,u,v,0)  + \int \frac{\hc^2}{s^2} \left ( 1 -\frac{1}{N^2} - \cA (\hc s)  \right ) \right \} , \\
I_{TD} &= I_{TD}^{(0)}-  \frac{4}{3 \cA (2 \hc)} \left \{ \int (u \alpha)^{-2}  \hat H'(s,u,v,\htheta) + \int \frac{\hc^2}{s^2} \left ( 1 -\frac{1}{N^2} - \cA (\hc s)  \right ) \right \} , 
\end{align}
where the function $\hat H'$ is obtained by subtracting the zero modes from $\hat H$ after Poisson resummation (see appendix~\ref{ap:largeN} for the details), and:
\begin{align}
&I_{TI}^{(0)}= -\frac{4}{3 \cA (2 \hc)} \int (u \alpha)^{-2}  \left (  \cA (2 \hc) -  \cA (\hc\alpha ) - \cA (\hc\alpha u /s ) + 1 - \frac{1}{N^2} \right ) \, , \label{eq:iapprox_p} \\
&I_{TD}^{(0)}=-\frac{4}{3N^2\cA(2\hc)}\sum_{n\ne0}\int(u \alpha)^{-2}e^{-\frac{\pi s n^2}{\hc N^2 \alpha u}}{\rm Re} \left\{\theta_3^2\left(0,\frac{i}{\hc\alpha}\right)\prod_{\mu}\theta_3\left(z_\mu,\frac{i}{\hc N^2\alpha}\right)-1\right\}, \label{eq:iapprox_np} 
\end{align}
where $z_\mu = \epsilon_{\mu \nu}n_\nu k/N + i v n_\mu /(\hc N^2 \alpha u )$, and where $n$ denotes a $d_t$-dimensional vector of integers taking values in the intervals $[-N/2, N/2)$ for even values of $N$, and $[-(N-1) /2, (N-1)/2]$ for odd ones. The leading correction to the infinite volume limit is in general driven by the contribution of $I_{TI}^{(0)}$ and $I_{TD}^{(0)}$, and depends on two quantities:  $\hc \alpha$ and $\hc \alpha u /s$.

\begin{figure}[t]
\centering
\includegraphics[width=0.85\linewidth]{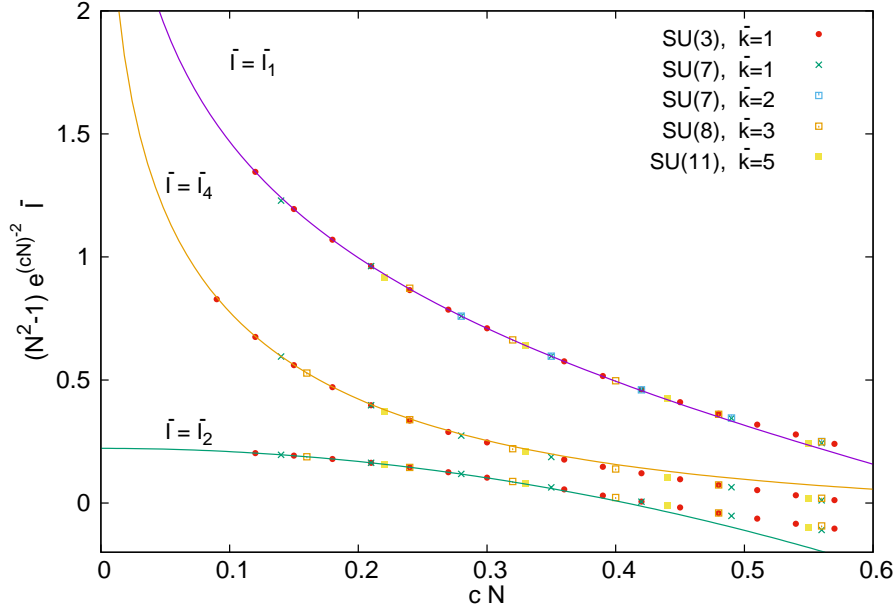}
\caption{The integrals $\tI_1$, $\tI_2$ and $\tI_4$ multiplied by the factor $(N^2-1) e^{(cN)^{-2}}$ as a function of $c N$. The continuous lines
are given by formulas~\eqref{eq:i1inf}-\eqref{eq:i4inf}.}
\label{f:cdep_exp}
\end{figure}

\begin{figure}[t]
\centering
\includegraphics[width=\linewidth]{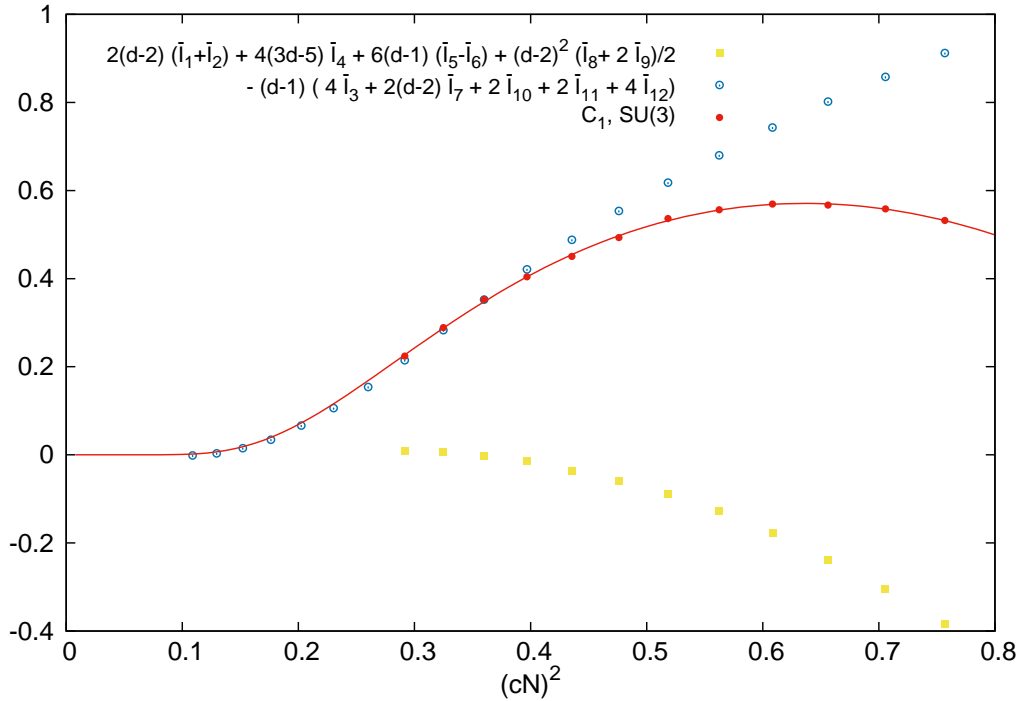}
\caption{Dependence on $cN$ of the SU(3) one-loop matching constant $\cC_1$. The continuous line is a fit to the functional form $f(cN)=e^{-(cN)^{-2}} (\alpha + \beta \log (cN) + \gamma cN + \delta c^2 N^2)$. We separated the contribution to $\cC_1$ into two pieces, plotted with open circles and squares.}
\label{f:c1_fit}
\end{figure}

\begin{figure}
\centering
\includegraphics[width=0.8\linewidth]{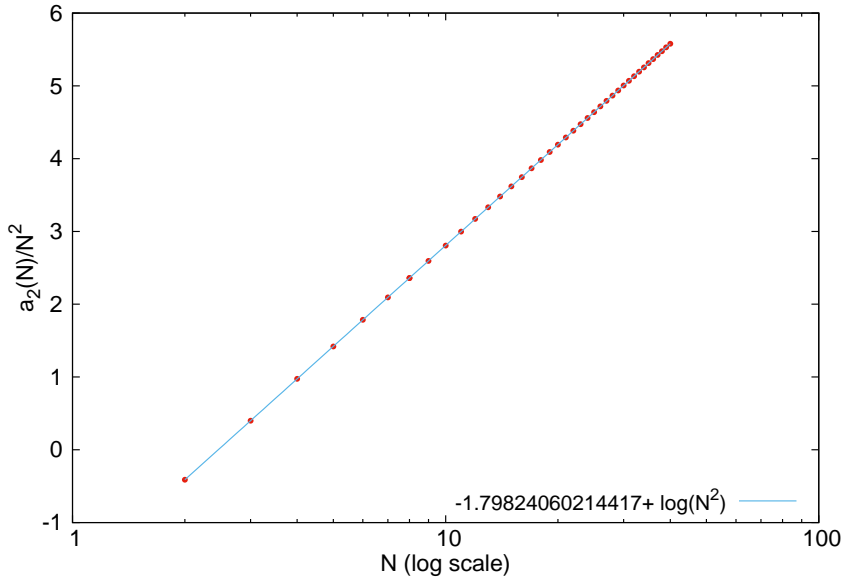}
\caption{Dependence on $N$ of the coefficient $a_2(N)/N^2$ appearing in eq.~\eqref{eq:a2}.}
\label{f:a2depN}
\end{figure}

The simplest case corresponds to integrals $\tI_1$, $\tI_2$ and $\tI_4$, for which both $\hc \alpha$ and $\hc \alpha u /s$ tend to zero in the $\hc \rightarrow 0$ limit in all of the integration range. The leading contribution, derived in appendix~\ref{ap:largeN}, is given by:
\begin{align}
\tI_1 &\rightarrow \frac{1}{9(N^2-1)}  e^{-(cN)^{-2}}   \left ( 1 + 3 \gamma_E - 3 \log \left (3 c^2 N^2\right) -3 c^2 N^2 \right ) ,
\label{eq:i1inf} \\
 \tI_2 &\rightarrow \frac{2}{9(N^2-1)}  e^{-(cN)^{-2}}    \left ( 1 -6 c^2 N^2 \right ) ,
\label{eq:i2inf} \\
 \tI_4 &\rightarrow \frac{1}{3(N^2-1)}  e^{-(cN)^{-2}}    \left ( -1 +  \gamma_E -  \log \left(9 c^2 N^2\right) + 3.544907702 \, c N - c^2 N^2 \right) .
\label{eq:i4inf}
\end{align}
Integrals for which the infinite volume contribution $I_i^\infty$ is UV-divergent at $d=4$, such as $\tI_1$ and $\tI_4$, have a leading correction that goes as $\sim \log (c^2 N^2) \exp (-1/(cN)^2)$. $I_2^\infty$ is UV-finite and the leading correction has a purely exponential decay in the thermodynamic limit, given by $\exp (-1/(cN)^2)$. We show in fig.~\ref{f:cdep_exp} the dependence of these integrals on $c N$ for several values of $\bar{k}$ and $N$, plotting their value multiplied by the factor $(N^2-1) \exp(1/(c N)^2)$. The continuous lines in the plot are given by the formulas presented above and describe very accurately the data for small $cN$. In the limit obtained by sending $N$ to infinity and $c$ to zero at small, fixed $cN$, the three integrals also go to zero with corrections of order $1/N^2$.

The general dependence of $\cC_1$ on $cN$ as $cN \rightarrow 0$ is in fact well described by a formula analogous to eq.~\eqref{eq:i4inf}; an example of this for the case of SU(3) is shown in fig.~\ref{f:c1_fit}, where $\cC_1$
is displayed as a function of $(cN)^2$. The continuous line in that plot is the result of a fit to the functional form $f(cN)=\exp (-1/(cN)^2) (\alpha + \beta \log (cN) + \gamma cN + \delta c^2 N^2)$. In order to push the calculation of $\cC_1$ to smaller values of $c$, we split it into two pieces, represented by the open blue circles and the yellow squares in the plot. The most relevant part comes from the contributions of $\tI_3$, $\tI_7$, $\tI_{10}$, $\tI_{11}$ and $\tI_{12}$, which we were able to compute down to values of $(cN)^2 \sim 0.1$. Asymptotically, this piece is described quite well by the function $f(cN)$, with a leading dependence on $c$ of the form $\log (cN) \exp (-1/(cN)^2) $. 

In the rest of this section, we will explore how the infinite volume limit is approached for the remaining integrals (excluding $\tI_1$, $\tI_2$, and $\tI_4$). The discussion is a bit more complex in their case, as the leading correction goes as $c^2$ for each of the integrals, but the corrections cancel out when all contributions to $\cC_1$ are considered. We will first analyze the case of $\tI_3$ in detail to see how the cancellation takes place, and then generalize it to all other cases. For this integral, in the $\hc\rightarrow 0$ limit, $\hc \alpha u /s$ goes to zero in the full integration range, and the leading dependence is given by:
\be
-\frac{4}{3 \cA (2 \hc)} \int_0^\infty dz  (3+2z)^{-2} \left \{ 1 -\frac{1}{N^2} -  \cA (\hc (3+2z) /2 ) \right \}
\, .
\ee
From this expression one can show (see app.~\ref{ap:largeN} for the details) that the dominant correction in the $cN \rightarrow 0$ limit is:
\be
I_0 = \frac{\pi (c N)^2}{6 N^2 \cA(2\hc) } \left(a_1 - \frac{1}{N^2} a_2(N)\right) + \cdots ,
\label{eq:a2}
\ee
with $a_1= -1.76508480122121275$ and, for instance, $a_2(N=3) = 3.59085631503990722$. The quantity $a_2(N)/N^2$ grows logarithmically with $N^2$, as shown in fig.~\ref{f:a2depN}. One can show that, in the infinite volume limit, all remaining integrals $\tI_i$ converge in the same manner, being proportional to $I_0$ with a proportionality coefficient of +1 for $i=5,6,7$, of -1 for $i=10,11,12$ and of 4 and -2 in the cases of $\tI_8$ and $\tI_9$ respectively. Combining eq.~\eqref{eq:order1} with these coefficients, it is easy to show that the total contribution of the leading $(cN)^2$ term to $\cC_1$ vanishes.

\begin{figure}[t]
\centering
\includegraphics[width=0.90\linewidth]{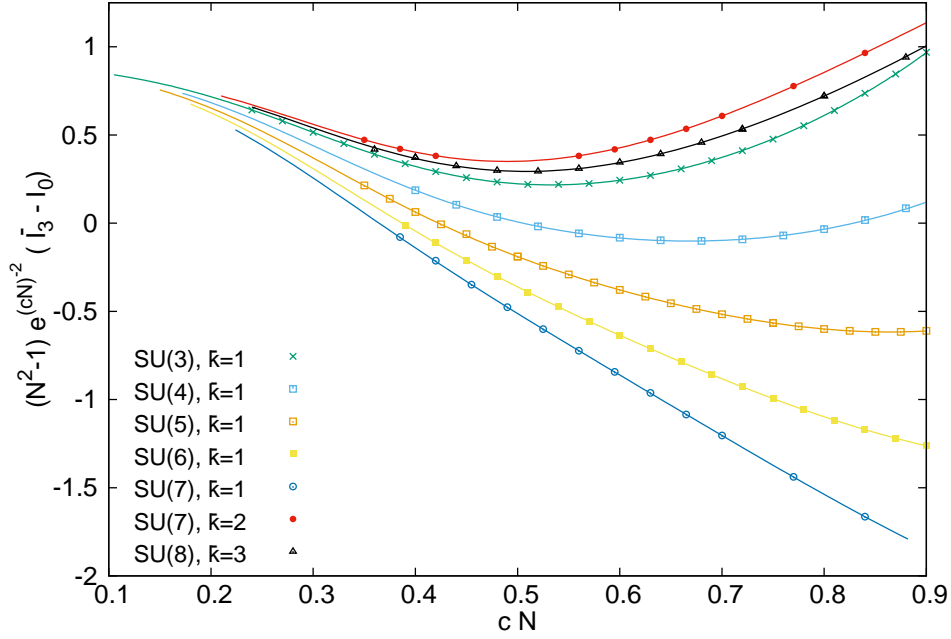}
\caption{The integral $\tI_3-I_0$ multiplied by the factor $(N^2-1) e^{(cN)^{-2}}$, plotted as a function of $cN$. The continuous line is obtained from the approximate expression given by the sum of eq.~\eqref{eq:iapprox_p} and ~\eqref{eq:iapprox_np}.}
\label{f:i3_exp}
\end{figure}

\begin{figure}[t]
\centering
\begin{subfigure}{.5\textwidth}
  \centering
\raggedleft
  \includegraphics[width=1.\linewidth]{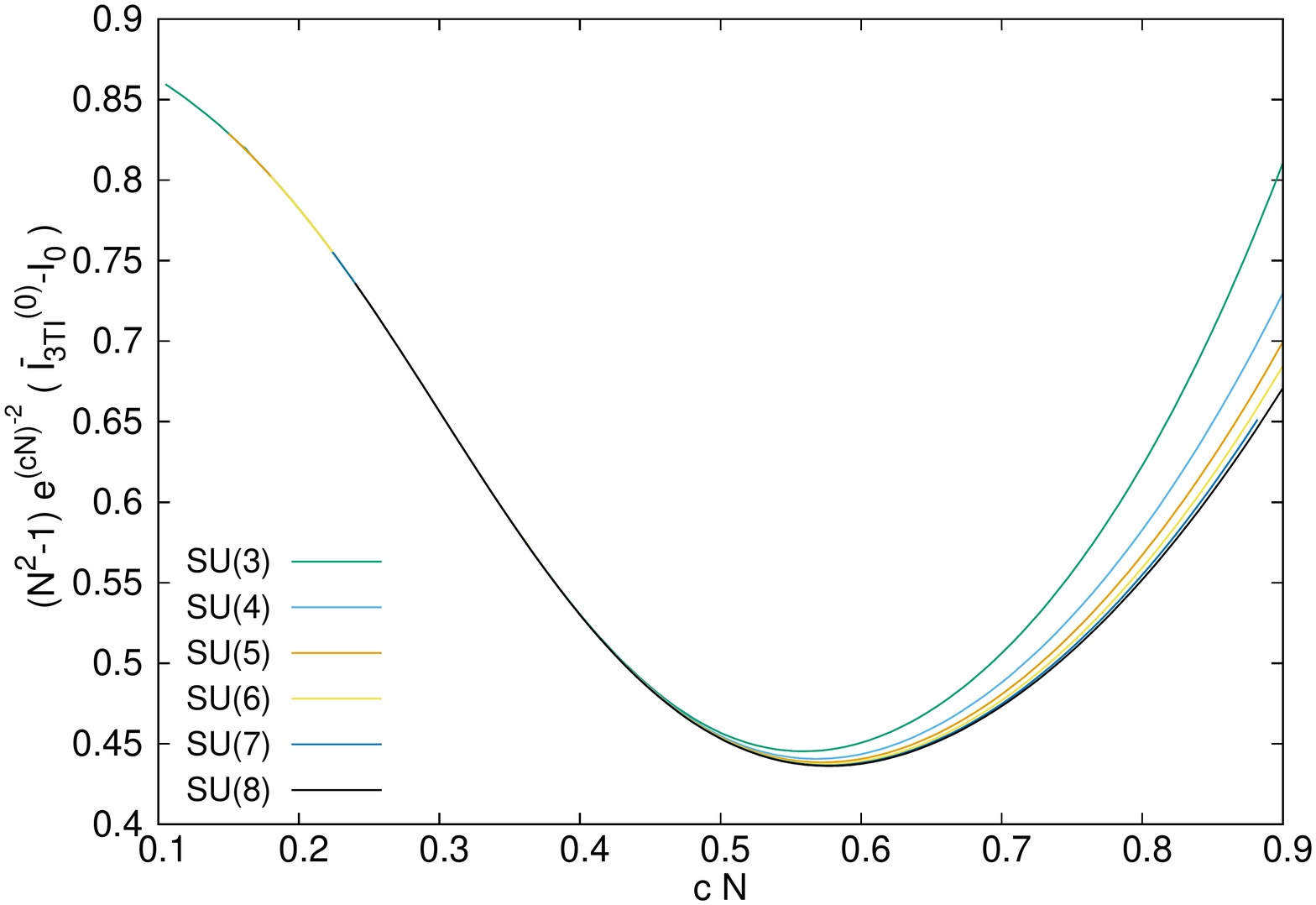}
\captionsetup{width=.95\textwidth}
  \caption{$\tI_{3 TI}^{(0)}-I_0$}
  \label{f:i3planar}
\end{subfigure}%
\begin{subfigure}{.5\textwidth}
  \centering
\raggedleft
  \includegraphics[width=1.\linewidth]{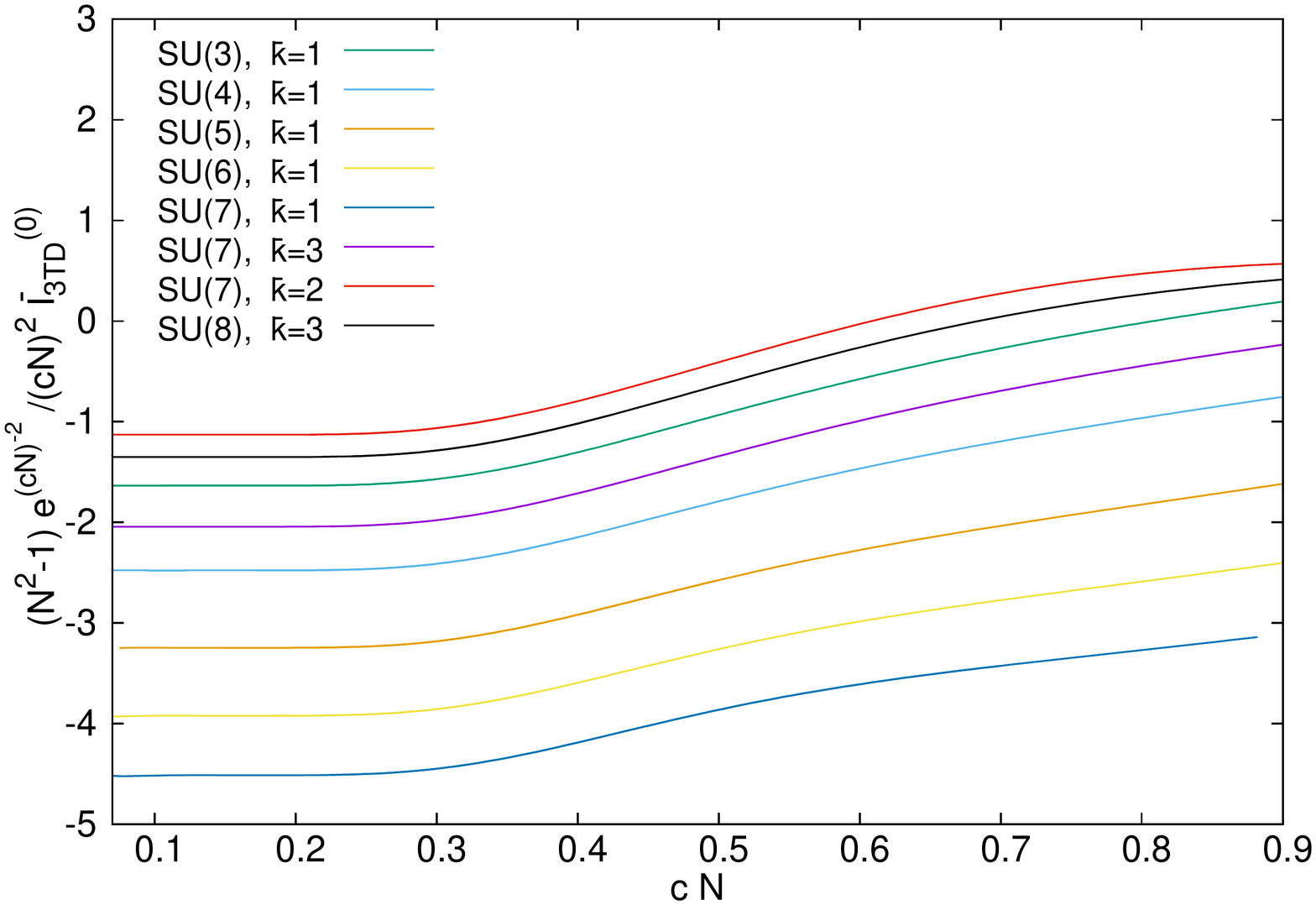}
\captionsetup{width=.95\textwidth}
  \caption{$\tI_{3TD}^{(0)}$}
  \label{f:i3nonplanar}
\end{subfigure}
\caption{The leading contributions to $\tI_3$ from eqs.~\eqref{eq:iapprox_p} and eq.~\eqref{eq:iapprox_np}, multiplied by factors $(N^2-1) e^{(cN)^{-2}}$ and $(N^2-1) e^{(cN)^{-2}}/(cN)^2$ respectively and displayed as a function of $cN$ for various values of $N$ and the magnetic flux $k$.}
\end{figure}

We did not analyze in detail how the different integrals approach zero after subtracting the quadratic piece in $c$, but, based on the results presented in fig.~\ref{f:c1_fit}, we expect other possible power like corrections to cancel out as well when combined to form $\cC_1$, the final result exponentially decaying towards zero with a leading dependence on $c$ of the form $\sim \log (cN) \exp \left (-(cN)^{-2}\right)/(N^2-1)$.   
A preliminary analysis was performed for the case of $\tI_3$, with the quantity $\tI_3-I_0$ times the factor $(N^2-1) \exp ((cN)^{-2})$ being shown in fig.~\ref{f:i3_exp} as a function of $cN$ for several values of $N$. Each point in that plot was obtained from the exact expression for $\tI_3$, and the continuous lines correspond to the approximate expression obtained combining eqs.~\eqref{eq:iapprox_p} and ~\eqref{eq:iapprox_np}. This decomposition is quite useful towards analyzing the $N$ dependence of the integral, and so we displayed each of the two pieces in figs.~\ref{f:i3planar} and ~\ref{f:i3nonplanar} as a function of the appropriate scaling variable $cN$. 

The $\htheta$-independent term is presented in  fig.~\ref{f:i3planar}, multiplied by the factor $(N^2-1) \exp ((cN)^{-2})$ scaling away most of its $N$ dependence. For $cN \rightarrow 0$, the integral decays exponentially as $\sim \exp (-(cN)^{-2})$, whereas in the large $N$ limit at fixed value of $cN$ it goes to zero with quadratic corrections in $1/N^2$. The analysis of the $\htheta$-dependent part is more complicated, as one needs to take into account the dependence on the magnetic flux $k$. The decay of $\tI_3$ towards zero is in this case faster than exponential, going as $\sim (cN)^2 \exp (-(cN)^{-2})$. This is shown in fig.~\ref{f:i3nonplanar}, where we plotted $\tI_{3 TD}^{(0)}$ multiplied by the inverse of this factor times $(N^2-1)$ as a function of $cN$ for various values of $N$ and the magnetic flux. In the large $N$ limit taken at fixed $cN$, this term also scales to zero as $1/N^2$. It would be interesting to study the $\htheta$-dependence for large values of $N$ in more detail for both this integral and the others, but such an analysis goes beyond the scope of this paper.

\section{Summary and conclusions}
\label{s:conclusions}

We computed the perturbative expansion at one-loop order of the $SU(N)$ twisted gradient flow coupling, including the matching to the $\MS$ infinite volume scheme at a renormalization scale $ \mu=1/(c\tilde{l})$ given by a combination of the size of the torus and the rank of the gauge group. The corresponding one-loop finite piece was determined numerically in the case of a two-dimensional non-trivial twist for which $\tl= l N$. 
The computation was done for a range of values of $c$ (the number relating the energy scale to the size of the torus), of the magnetic flux, and for several values of the rank $N$ of the gauge group, allowing us to obtain the ratio of $\Lambda$ parameters between the TGF scheme and the $\MS$ one.

Moreover, we deemed it interesting to explore the dependence of the coupling on the number of colors and the magnetic flux in a bit more depth, and so we analyzed the dependence of $\lambda_{TGF}$ in two different limits. First, we studied the limit in which $N$ and the torus size are sent to infinity and zero respectively in such a way as to keep $\tl$, and hence the renormalized 't Hooft coupling at scale $ \mu=1/(c\tilde{l})$, fixed. This is a singular large $N$ limit in the spirit of those introduced in \cite{AlvarezGaume:2001tv}, albeit a rather non-standard one since non-planar, $\theta$-dependent diagrams survive the limit as long as $\tl$ is finite.  The connection of this case to non-commutative Yang-Mills theory is straightforward through the use of the Morita duality: the non-commutative dual torus is of length $\tl$ and has a dimensionless non-commutativity parameter given by $\htheta=\bar k/N$. Our analysis also supports the observation, first presented in \cite{Chamizo:2016msz,Perez:2018afi}, that the avoidance of tachyonic instablities when taking the singular limit is only possible for a zero-measure, though uncountable, set of values of $\htheta$. Curiously, one of the successful cases, of limiting parameter $\htheta= (3-\sqrt 5)/2$, relies on a sequence of Fibonacci numbers with $k=F_{i-2}$ and $N=F_i$ with $F_i$ denoting the $i$-th element of the Fibonacci series~\cite{Chamizo:2016msz}. 

The second limit at which we looked was the thermodynamic limit, in which $c$ is sent to zero and $\tilde{l}$ is sent to infinity while keeping the energy scale $ \mu$ constant. This leads to the one-loop expression of the 't Hooft gradient flow coupling at infinite volume~\cite{Luscher:2010iy}. Our results give support to the reduction idea, in the sense that the $SU(\infty)$ coupling in the thermodynamic limit can also be recovered at fixed torus size by sending $N$, and hence $\tl$, to infinity, in which case the limit is approached with $1/N^2$ corrections.

\section*{Acknowledgments}

We would like to thank Antonio Gonz\'alez-Arroyo and Alberto Ramos for many valuable discussions on both this topic and related ones. We acknowledge financial support from the MINECO/FEDER grant FPA2015-68541-P and the MINECO Centro de Excelencia Severo Ochoa Programs SEV-2012-0249 and SEV-2016-0597. E.I. Bribi\'an acknowledges support under the FPI grant BES-2015-071791. The numerical computations presented in this paper have been carried out at the IFT Hydra cluster and with computer resources provided by CESGA (Galicia Supercomputing Center).


\appendix 
\section{The Feynman rules with twisted boundary conditions}
\label{ap:frules}

The Feynman rules for the set of irreducible twist tensors used in this work have been derived in various contexts both in the continuum (see for instance~\cite{Perez:2014sqa} and references therein for a review) and in the lattice regularized version of the theory~\cite{GonzalezArroyo:1982hz,Luscher:1985wf,Snippe:1997ru,Perez:2017jyq}. In this appendix, we will summarize the ones relevant to our work, derived in the continuum.

The set of allowed gauge transformations in our theory will be restricted to those preserving the form of the boundary conditions in eqs.~\eqref{eq:tbc},~\eqref{eq:pbc}, using the irreducible twist given in eq.~\eqref{eq:twist}, and the remaining gauge degrees of freedom will be fixed using a generalized covariant gauge of parameter $\xi$ consistent with the boundary conditions. After scaling the gauge potential with the bare coupling $g_0$, the Lagrangian density, including the gauge fixing terms, reads:
\be
{\cal L} = \half \Tr ( F_{\mu \nu}^2 )+ {\frac{1}{\xi}} \Tr (\partial_\mu A_\mu )^2
- 2 \Tr (\bar c \, \partial_\mu D_\mu c)\quad ,
\ee
where  $D_\mu \equiv \partial_\mu + i g_0 A_\mu$ is the covariant derivative and $c$,
$\bar c$ denote the ghost fields.

One may then obtain the propagators of the gauge and ghost fields using the Fourier expansion of the gauge potential given in eq.~\eqref{eq:fourier}, along with an analogous one for the ghost fields:
\begin{align}
&P_{\mu \nu} (\p, \q) = {\frac{1}{\p^2}} \Big (\delta_{\mu \nu} - (1-\xi) \ {\frac{\p_\mu \p_\nu} {\p^2}}\Big) \  \delta(q+p) \quad , \label{eq:gaugeprop} \\
&P_g(\p,\q) =   {\frac{1}{\p^2}}  \delta(\q+\p)\quad \, , \label{eq:ghostprop}
\end{align}
where the momenta appearing in these expressions are quantized in units of the effective size $\tl$.

The Feynman rules for the vertices are then obtained from the commutation relations in eq.~\eqref{eq:comm}, and are expressed in terms of the momentum-dependent structure constants $F(p,q,-q-r)$. The terms contributing to minus the gauge fixed action are the following:
\begin{itemize}
\item
3-gluon term:
$$ \frac{1}{3!} \cV^{(3)}_{\mu_1 \mu_2 \mu_3}(p_1, p_2,p_3) \left (\prod_{i=1}^3 A_{\mu_i} (p_i)\right )  \, ,$$
with:
\bal
\cV^{(3)}_{\mu_1 \mu_2 \mu_3}(p_1, p_2,p_3)&= i g_0 V^{-\half}
  F(p_1, p_2,p_3)           \, \delta \Big (\sum_{i=1}^3 \p_i\Big )\times \\
\Big (( p_3-p_2)_{\mu_1} \delta_{\mu_2 \mu_3}
   &+  ( p_1-p_3)_{\mu_2} \delta_{\mu_1 \mu_3}
   +  ( p_2-p_1)_{\mu_3} \delta_{\mu_1 \mu_2}
\Big ) \, . \nonumber
\end{align}
\item
4-gluon term:
$$ \frac{1}{4!}  \cV^{(4)}_{\mu_1 \mu_2 \mu_3\mu_4}(p_1, p_2,p_3,p_4) \left (\prod_{i=1}^4 A_{\mu_i} (p_i)\right)  \, ,$$
with:
\bal
& \cV^{(4)}_{\mu_1 \mu_2 \mu_3 \mu_4} (p_1, p_2,p_3,p_4) = - g_0^2   V^{-1}  \, \delta \left(\sum_{i=1}^4  \p_i\right) \times  \\
\Big( &F(p_1,p_2,-p_1-p_2) F(p_3,p_4,-p_3-p_4) (\delta_{\mu_1 \mu_3} \delta_{\mu_2 \mu_4}-\delta_{\mu_2 \mu_3} \delta_{\mu_1 \mu_4})  \nonumber\\
+&F(p_2,p_3,-p_2-p_3) F(p_4,p_1,-p_4-p_1)
(\delta_{\mu_2 \mu_4} \delta_{\mu_3 \mu_1}-\delta_{\mu_3 \mu_4}
\delta_{\mu_2 \mu_1})  \nonumber\\
+&F(p_1,p_3,-p_1-p_3)  F(p_2,p_4,-p_2-p_4)
(\delta_{\mu_1 \mu_2} \delta_{\mu_3 \mu_4}-\delta_{\mu_3 \mu_2}
\delta_{\mu_1 \mu_4}) \Big )\nonumber \, .
\end{align}
\item
Ghost-gluon term:
\be
\cV^{(gh)} =   i  g_0 V^{-\half}  F(p_1,p_2,p_3) \
\, p_{1\mu} \  \bar c (p_1) A_\mu (p_2)
c(p_3)  \, \delta \left (\sum_{i=1}^3 p_i\right ) \,  .
\ee
\end{itemize}

These rules can be easily used to derive different quantities, such as the one-loop correction to the propagator. At order $\mathcal{O}(g_0^2)$ and in the Feynman gauge ($\xi=1$), the vacuum polarization tensor can be obtained as shown in ref.~\cite{Perez:2014sqa}:
\bal
\Pi_{\mu \nu} (p) &=
\half g_0^2   V^{-1}  \sum_{q}
F^2(p,q,-p-q) \, \frac{1}{q^2 (p+q)^2} \, \times
\label{eq.vacpol}
\\
&\Big\{ 4 \, \left (\delta_{\mu \nu} p^2 -p_\mu p_\nu\right ) + (d-2)  \, \left ((p_\mu + 2 q_\mu)
(p_\nu  + 2 q_\nu) -  2 \delta _{\mu \nu} q^2 \right ) \Big \}
\nonumber \, .
\end{align}

\section{Integral form of the energy density at NLO}
\label{ap:intform}

As we recall, the energy density at NLO in the twisted gradient flow scheme can be expressed in terms of several integrals. In section ~\ref{s:integral}, we chose for both clarity and concision to show a single example of the derivation of these integrals, and left the expression of the full seven $\mathcal{O}\left(\lambda_{0}^{2}\right)$ contributions to the observable $\left\langle E/N\right\rangle$ in terms of the integrals for this appendix. The contributing $\cE_{i}$ terms are the following:
\begin{align}
&\cE_{0}^{\left(1\right)}=  \half \lambda_0^2 \left\{ \left(3d-2\right)\mathbb{I}_{1}+2 \left(d-2\right)^{2} I_9\right\} 
\label{eq:e0b},\\
&\cE_{1}=  -3 \lambda_{0}^{2}\left(d-1\right)\mathbb{I}_{2}
\label{eq:e1b},\\
&\cE_{2}=  \lambda_{0}^{2}\left\{ 8\left(d-1\right)I_5+d\left(\mathbb{I}_{2}-\mathbb{I}_{1}\right)+4\left(d-2\right)\mathbb{I}_{3}\right\} 
\label{eq:e2b},\\
&\cE_{3}=  \half \lambda_{0}^{2}\left\{ \left(d-2\right)\left(4 I_1+4 I_2 +2\mathbb{I}_{2}-\mathbb{I}_{1}
- I_8\right)-4\left(d-1\right)\left(I_{10}+I_{11} + 2 I_3 -2 I_6 \right)\right\} 
\label{eq:e3b},\\
&\cE_{4}=  \half \lambda_{0}^{2} \,  d\left(d-1\right)I_8 
\label{eq:e4b},\\
&\cE_{5}=  -\lambda_{0}^{2}\left(d-1\right)\left(10 I_6+I_8-\mathbb{I}_{2}\right) 
\label{eq:e5b},\\
&\cE_{6}=  -2 \lambda_{0}^{2}
\left\{ \left(d-1\right) (I_5+2 I_{12}) +2\left(d-2\right)\mathbb{I}_{3} +\left(d-2\right)\left(d-1\right)I_7 - 2\left(3d-5\right)I_4 
\right\} .
\label{eq:e6b}
\end{align}
When summing all of the terms contributing to $\cE^{(1)}(t)$, the $\mathbb{I}_{i}$ terms cancel out, and thus:
\begin{align}
\cE^{(1)}(t)=& 2(d-2)\left(I_{1}+I_{2}\right)-4(d-1)I_{3}+4(3d-5)I_{4}+6(d-1)\left(I_{5}-I_{6}\right)
\label{eq:ENLO}
\\
-& 2(d-2)(d-1)I_{7}+ \half (d-2)^2  (  I_{8}+ 2 I_{9} ) -2(d-1) \left(I_{10}+I_{11}\right)-4(d-1)I_{12}
\, ,
\nonumber
\end{align}
which is the NLO result given in eq.~\eqref{eq:order1}. The intervening $I_i$ integrals are given in eqs.~\eqref{eq:i1} -~\eqref{eq:i12}, in terms of the auxiliary $\Phi(s,u,v,\hat{\theta})$ functions from ~\eqref{eq:calN}.

\section{Regularization of \texorpdfstring{$I_9$}{I9}}
\label{ap:I9}
 
We will present here the full details of the procedure to regularize the integral $I_9$, defined in eq.~\eqref{eq:i9}, which differs slightly from the general treatment described in sec.~\ref{s:regularization}. As we recall, the initial integral is split into three terms:
\be
I_9 (t') =  I_9(\Phi-\theta(1-z) \Phio, t')  - I_9(\theta(z-1) \Phio, t') +  I_9(\Phio, t') \, ,
\ee
with the Heaviside function $\theta$ restricting the integration intervals in $z$.
The first term on the r.h.s. of this expression is already finite in four dimensions, whereas the other two will be shown to be so as well after analytical continuation to $d=4$.

Let us start by discussing the treatment of $I_9(\Phio, t')$. The original integral can be rewritten as:
\be
 I_9( \Phio, t')  = \half \int_{0}^{\infty} dz\int_{0}^{\infty}dy\int_{0}^{1}dx\,  (x^2 z \partial_{t'} - d) \Phi^{(0)}\left(2t'+xz+y ,z,xz\right)
\, ,
\label{eq:api9}
\ee
which, after some algebraic manipulation, becomes:
\be
I_9( \Phio, t')  =
\frac{{\cal N} \hc^{-d}}{d-2} \,  {\cal I} \int_{0}^{\infty}dz\, z^{1-d/2}  (2 t' + z )^{-d/2}
 {\cA}\left (\hc (2t'+z)\right)
\, ,
\ee
where
\be
 {\cal I} \equiv \int_{0}^{1}dx\, \left(x (1-x) \right)^{d/2-2} \left (d x (1-x) - (d-2) x^2 \right)\, .
\ee
This integral ${\cal I}$ is in $d=4-2\epsilon$ dimensions of order $\epsilon$ times zero.
The asymptotic behavior at $z=0$ is then obtained by expanding ${\cA}(\hc (2t'+z))$ around $z=0$, leading to:
\be
I_9( \Phio, t') = {\cA}(2\hc t') I_9^{\infty} (t')\, ,
\ee
where
\be
I_{9}^\infty (t')= \frac{{\cal N} \hc^{-d}}{d-2}\,   {\cal I} \int_{0}^{\infty} dz \, z^{1-d/2} (2 t' + z )^{-d/2}
\, .
\ee
The integral over $z$ presents a pole in $1/\epsilon$, but it is cancelled when multiplied by ${\cal I}$, leading to a final result that is identically zero for $d=4$.
This $I_{9}^\infty$ is precisely the integral appearing in the infinite volume calculation, and vanishes, as we have just seen, 
for $d=4$ in dimensional regularization.

The remaining $I_9(\theta(z-1)\Phio, t')$ term can be treated in a similar way. One takes the initial expression:
\be
I_{9}(\theta(z-1) \Phio,t')=  \int_{1}^{\infty}z dz\int_{0}^{\infty}dy\int_{0}^{1}dx\, \partial_{z'}\Phio\left(2t'+x z+y,z',xz\right)\Big|_{z'=z}\, ,
\ee
and rewrites it in a form identical to eq.~\eqref{eq:api9}, only with the integral over $z$ restricted to the interval $[1,\infty]$. After some manipulation, 
the regularized result becomes:
\begin{align}
&I_9^{\rm reg}(\theta(z-1)\Phio, t') = -\frac{ {\cal N} \hc^{-d}}{d-2} \int_{0}^{\infty}dz\, \left \{
 2 (2t'+z)^{-d/2}  \cA\left(\hc(2t'+z)\right) \right .\\
&\left .+ (d-2)
 \int_{0}^{1}dx\, \left(2t'+x(1-x)+z\right)^{-d/2} \cA\left (\hc(2t'+x(1-x)+z)\right) 
  \right \}
\, ,
\nonumber
\end{align}
which is finite in $d=4$ dimensions, and which we were able to evaluate numerically.

\section{Numerical implementation of the integration algorithm}
\label{ap:numerical}

In order to perform the numerical computation required to obtain the results presented in sec.~\ref{s:results}, we prepared a code in C++ to compute the values of the $\Phi(s,u,v,\hat{\theta})$ functions and their derivatives at any point, and integrate them along the corresponding ranges using the trapezoidal rule up to a target precision. We will in this section begin by explaining how the computation of each $\Phi(s,u,v)$ is performed, and then detail how the integration algorithm works.

\subsection{Momentum Sums}

As we recall, we had to compute the following quantity:
\be
\Phi^\rf (s,u,v)= H\left(s,u,v,0\right)-H\left(s,u,v,\htheta\right) - \Phi^{(0)} (s,u,v) \, ,
\ee
which was made finite through the procedure explained in sec.~\ref{s:regularization}, and where $H$ and $\Phi^{(0)}$ were defined as in that very section, taking $d=4$. The three r.h.s. terms can be rewritten in terms of momentum sums of the general form:
\begin{equation}
\underset{M \in \mathbb{Z}^s}{\sum}\exp\left(-\pi M^{t}XM\right)=\left(\det X\right)^{-\frac{s}{2}}\underset{M \in \mathbb{Z}^s}{\sum}\exp\left(-\pi M^{t}X^{-1}M\right), \quad s\in \mathbb{Z},
\end{equation}
where we introduced a generic matrix $X$ to denote either $A_0$ or $B$ from eqs.~\eqref{eq:a0def} and~\eqref{eq:bdef}. We used Poisson resummation to write the sums in terms of both $X$ and its inverse, allowing us to simultaneously compute several equivalent versions of the three terms of $\Phi^\rf$, which let us exploit the fact that convergence speed depends on the $(s,u,v)$ point being considered to speed up the program. We defined eight quantities to be computed:
\begin{align}
&E_{0}=\underset{m\neq 0\text{ mod }N}{\sum}e^{-\pi M^{t}B_{\theta}M},
&E_{1}=\underset{m,n}{\sum}e^{-\pi M^{t}B_{0}M},
\\
&E_{2}={(\det B_{\theta})}^{-\frac{1}{2}}\underset{m,n}{\sum}e^{-\pi M^{t}B_{\theta}^{-1}M},
&E_{3}={(\det B_0)}^{-\frac{1}{2}}\underset{n\neq 0}{\sum}e^{-\pi M^{t}B_{0}^{-1}M},
\\
&E_{4}=\underset{m=0\text{ mod }N}{\sum}e^{-\pi M^{t}B_{\theta}M},
&E_{5}={(\det B_0)}^{-\frac{1}{2}}\underset{n=0}{\sum}e^{-\pi M^{t}B_{0}^{-1}M},
\\
&E_{6}={(\det \tilde{B})}^{-\frac{1}{2}}\underset{n=0}{\sum}e^{-\pi M^{t}\tilde{B}^{-1}M},
&E_{7}=\underset{n=0}{\sum}e^{-\pi M^{t}\tilde{B}M},
\end{align}
where we used the shorthands $B_\theta \equiv B(\hc s,\hc u,\hc v,\hat{\theta})$, $B_0 \equiv B(\hc s,\hc u,\hc v,0)$, and $\tilde{B} \equiv B(\hc sN^2,\hc u,\hc vN,0)$ for clarity. Several of these expressions are redundant:
\be
E_{1}=E_{3}+E_{5},\qquad E_{2}=E_{0}+E_{4},\qquad E_{6}=E_{7},
\ee
allowing us to rewrite the observable $\Phi^\rf (s,u,v)$ in four equivalent forms:
\begin{align}
\Phi^\rf \left(s,u,v\right) &= \mathcal{N} \left[ E_1(E_1-E_0-E_4)+E_5(E_6-E_5)\right], \\
\Phi^\rf \left(s,u,v\right) &= \mathcal{N} \left[E_1(E_1-E_0-E_4)+E_5(E_7-E_5)\right], \\
\Phi^\rf \left(s,u,v\right) &= \mathcal{N} \left[E_5(E_6-E_2)+(E_3+E_5)(E_3+E_5-E_2)\right], \\
\Phi^\rf \left(s,u,v\right) &= \mathcal{N} \left[E_5(E_7-E_2)+(E_3+E_5)(E_3+E_5-E_2)\right] .
\end{align}
We derivated these four equivalent expressions in the integrals in which it was required, simply using the chain rule and computing the derivatives of each $E_{i}$ function when needed.

We will skim over the details of the algorithm used to generate the momenta in the sums, simply mentioning that we defined a four-dimensional integer vector $M^{t}=(m_{1},n_{1},m_{2},n_{2})$ and generated the corresponding combinations of integers $m_i,n_i$, using the $M\rightarrow-M$ symmetry in the integrand to shorten the computation time. The momentum tetrads were generated in an orderly manner, starting with all contributions of the tetrads with $|m_i|,|n_i|=0,1$, then adding the ones with some $|m_i|,|n_i|=2$, then $|m_i|,|n_i|=3$ and so on and so forth, adding terms with momenta of increasing order until the sum converges (in the sense that we will detail below).

Thus, the code simply runs through momentum tetrads of increasing order, and passes them through a filter that checks whether or not $m$ is proportional to $N$ and whether or not $n=0$, computing and adding the relevant exponential terms to each of the eight $E_{i}$ terms. Once every tetrad of a particular order has been processed, the program computes the value of $\Phi^\rf$ up to that order in the four equivalent ways shown earlier, and checks whether the variation of each term between the previous order and the new one is smaller than a set quantity $\epsilon$ times the value of the function. If that turns out to be the case for any of the four expressions, the sum is considered to have converged and that particular $\Phi^\rf$ is returned as the result. To avoid early spurious convergences, we set a minimum order of four for the sum. The same relative error $\epsilon$ was also used as the convergence criterion for the integration algorithm (see next subsection), and ranged between $10^{-3}$ and $10^{-8}$ depending on the integral, due to differences in runtime between them.

\subsection{Integration Algorithm}

Now that we explained how the integrand is computed at each point, we may focus on the integration algorithm, for which we chose to use a fairly standard trapezoidal rule for multiple integrals in which the integral along each coordinate is approximated using an increasing number of trapezoids until a target precision is reached. We will begin by quickly illustrating how a generic single-dimensional integral works in our code, generalize it to the multiple ones, and then mention a few specific choices of strategy.

Consider thus a single integral over a finite interval, say for instance the interval $z \in [0,1]$. The code begins by computing the value of the integrand, which in this case would be the $\Phi$ function, at the beginning and end of the interval, and approximates the integral with a trapezoid. The integrand is then determined at the middle point $z=0.5$, and the integral is approximated with the two $z \in [0,0.5],[0.5,1]$ trapezoids. Then, at third order, the integrand is obtained in the midpoints of the previous trapezoids, and the integral is approximated with the four trapezoids $[0,0.25],[0.25,0.5],[0.5,0.75],[0.75,1]$. This subdivision generated by computing the integrand at the midpoints goes on, until the variation in the approximated integral between one order and the next is smaller than a set target $\epsilon$ (the same that we used for the $\Phi$ functions above) times the value of the integral at that order, at which point we consider that convergence has been reached and the integral is finished. As we mentioned earlier, in our runs $\epsilon$ ranged between $10^{-3}$ and $10^{-8}$.

Multiple integrals are trivial in such a setting: one simply starts with the integral over the outermost coordinate, $z$, but at every point in which the integrand needs to be determined instead of computing the $\Phi$ function, one recursively calls the integration routine to obtain the integral over the next coordinate.

To allow for easier parallelization, and since the integrand tends to have more structure near $z=0$, we chose to split the integral in $z$ into a set of pre-chosen subintervals, with a shorter step size at smaller values of $z$, and treated the integration along each of these subintervals separately. To avoid spurious convergences, we imposed a minimum of eight points in each integration subinterval. Moreover, in the cases in which the integrals went up to infinity in the $z$ coordinate, we ran the integration code up to $z_{max}=10^{4}$ and extrapolated the result by fitting the results of the last ten subintervals to a simple shifted exponential of the form $I_{fit} = a_0 - a_1 e ^ { - a_2 (z - z_{max}) }$, using the fitted $a_0$ as the final result of the integral. A simple least squares method algorithm was used to perform the fits.

There were a couple of peculiarities worth mentioning regarding integrals $I_8$ and $I_9$. For the former, and after performing a change of variables so that the second integral runs up to $x=1$, we noticed that the contribution to the integral is concentrated around $z=0$, with the profiles of the integrand over $x$ peaking at small values of $z$ and vanishing after a range $\sim z^{-1}$. This means that the strategy to keep dividing the integration interval into halves in the $x$ coordinate is quite inefficient, as the contribution is concentrated in a small region and one is throwing many points into areas that are effectively zero. To avoid this issue, we chose to subdivide the inner integral into 1,5,50,500 and 1000 equal subintervals as $z$ runs up to 1,10,1000 and 10000 respectively. As soon as the integral over two consecutive subintervals in the $x$ axis vanishes for $z>1$, the subintervals that follow are ignored entirely, greatly speeding up the computation without affecting the result.

The case of $I_9$ is a bit special in that the regularization was different from the other integrals, with a Heaviside $\theta(1-z)$ function being introduced in the integrand (see the end of sec.~\ref{s:regularization} and app.~\ref{ap:I9} for the specifics) and separating the bits before and after $z=1$. For the numerical computation, we performed the same change of integral as in $I_8$ to make the second integral run up to $y'=1$, but then the Heaviside function became a $\theta(1-y'z')$ function, with the integrands being different before and after this point. As convergence turned out to be painfully slow when both integrands were considered jointly, we simply forced the integrals in $y'$ to be split from $z'=1$ onwards into two subintervals $[0,1/z']$ and $[1/z',1]$, with the convergence of each side being considered separately.

Due to the procedure we used to determine the convergence of the integrals, for a given integral $\mathcal{I}$, and dubbing the number of integrals to perform $n_{i}$ (single, double or triple), the final error of the integral is:
\begin{equation}
\Delta\mathcal{I}=(1+n_{i})\epsilon\mathcal{I}.
\end{equation}
This comes from the fact that both the error of the $\Phi$ functions and the convergence criterion for the integrals is given by the same $\epsilon$, so for a single integral:
\begin{equation}
\mathcal{I}+\Delta\mathcal{I}=\left(1+\epsilon\right)\sum\left(\Phi+\Delta\Phi\right)\simeq\left(1+2\epsilon\right)\mathcal{I}.
\end{equation}

Additional integrals simply add extra $1+\epsilon$ factors, which end up generating the $(1+n_{i})$ term. In the cases where the integrals ran up to infinity in $z$ and had to be fitted, we presented as the final error either $\Delta\mathcal{I}$ or the error from the fit itself, whichever was larger.

Moreover, some issues were caused by some computed quantities hitting machine precision, slowing down the computation while leaving the results effectively unaffected. To deal with them, we introduced several hard cuts in the integrals, integrands and determinants. In particular, we made it so that any $\Phi$ function returning a value under $10^{-12}$, any inner integral returning any value under $5\times10^{-12}$ (or $10^{-10}$ in the cases of a few intervals in which using $5\times10^{-12}$ led to severe slowdowns), and any exponential returning a result over $10^{-13}$ is automatically set to be exactly zero. The cut in the integrals is also used in the convergence checks we mentioned earlier: whenever the value of the integral times $\epsilon$ becomes smaller than the precision cut, the precision cut is used as the convergence criterion instead.

\begin{figure}[ht]
\centering
\includegraphics[width=0.50\linewidth]{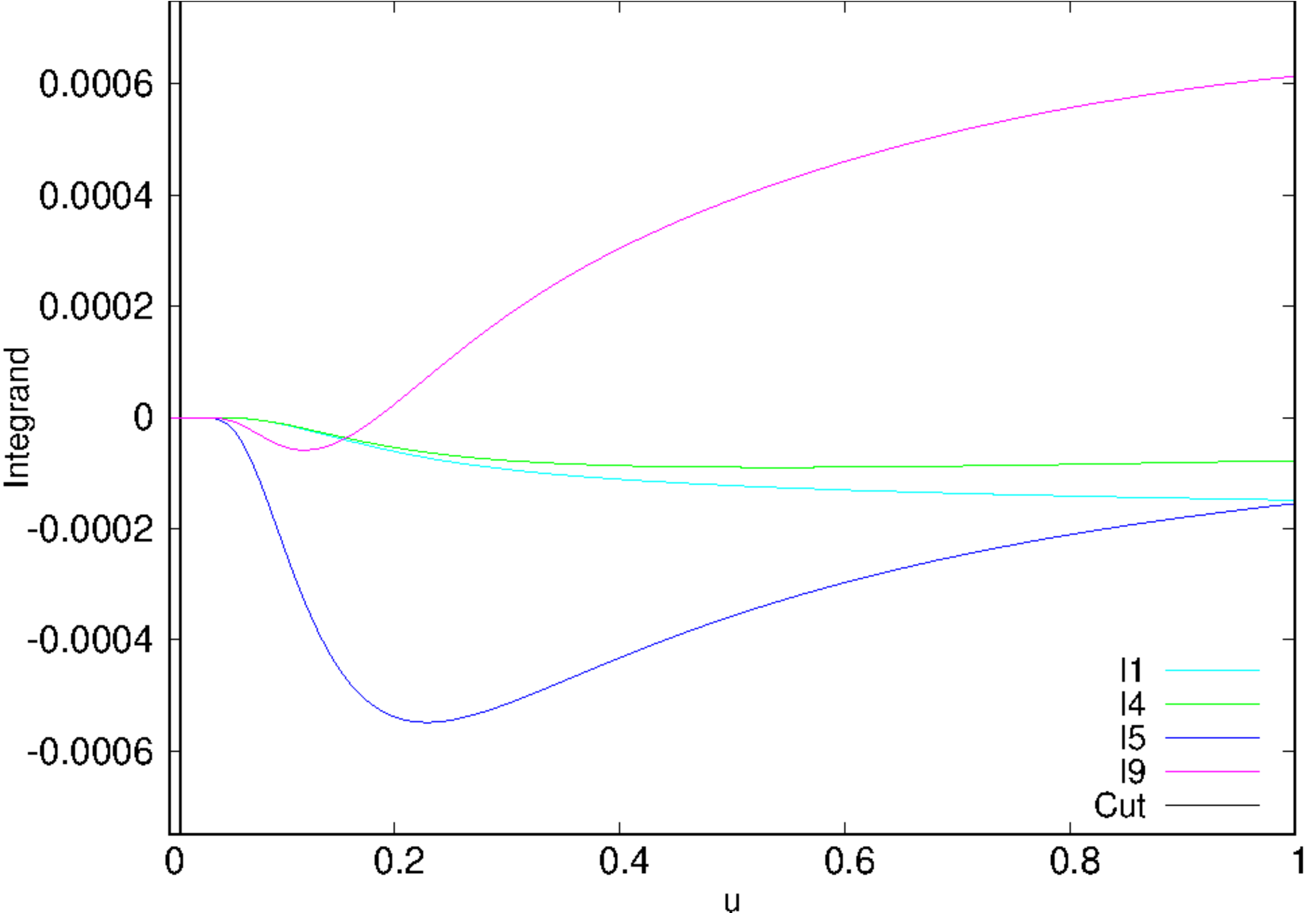}
\caption{ We display several examples of the profile of the integrand as a function of $u$ near $(s,u,v)=(2,0,0)$ for several integrals for $c=0.7$, to illustrate that the cut introduced in $u$ (displayed as a vertical line near the origin) has no effect on the resuts.}
\label{f:u_cut}
\end{figure}

Lastly, we need to mention that, despite the integrals computed being finite, convergence near the point $(s,u,v)=(2,0,0)$ can become quite slow, as the integrand approaches machine precision. To address this issue, a cut in $u$ was introduced, setting the integrand to zero when $u<0.01$ in the integrals in which such point is part of the integration region (namely, in $I_i$ for $i=1,4,5,7,9,11,12$). This cut does not appreciably change the results, as the contribution of the excluded area is well below the uncertainty of the total result. To illustrate this, we show in fig.~\ref{f:u_cut} some examples of the profile of the integrand near the aformentioned $(s,u,v)=(2,0,0)$ point, in which one can both see that the integrand is indeed finite and that the area excluded by the cut is negligible compared to the rest of the integrand.

\section{The infinite volume and large \texorpdfstring{$N$}{N} limits}
\label{ap:largeN}

In this appendix, we will derive the formulas mentioned in sec.~\ref{s:colours}, which were used to analyze the $N$ and $\htheta$ dependence of $\cC_1$ at NLO in the coupling for the case of a two dimensional twist ($d_t=2$). As we recall, the contributions to $\cC_1$, barring the one from $I_9$ which is slightly different, can be written in the form:
\be
\tI= \frac{4}{3 \cA (2 \hc)} \int (u \alpha)^{-2} \left(\hat H(s,u,v,0) - \hat H(s,u,v,\htheta) - \cA(2\hc)\right)
\, ,
\label{eq:idef}
\ee
where the notation $\int$ was used to refer generically to the integrals from eqs.~\eqref{eq:i1}-\eqref{eq:i12}, including the prefactors multiplying the $\Phi$ function and derivatives when required. The function $\hat H(s,u,v,\htheta)$ was defined through the relation:
\be
H(s,u,v,\htheta) = \Phi^\infty(s,u,v) \hat H(s,u,v,\htheta)
\, ,
\ee
where $\Phi^\infty$ was defined in eq.~\eqref{eq:phinfty} and $H$ is the function given in eq.~\eqref{eq:Hdef}, in terms of which we rewrote the $\Phi$ functions entering the integrals:
\be
\Phi(s,u,v,\htheta)= H(s,u,v,0)-H(s,u,v,\htheta)
\, .
\ee
The function $\hat H$ can be expressed as:
\be
 \hat H(s,u,v,\htheta) ={\rm Re} \left \{F_1(\alpha, u, v,0,4-d_t )\left (  F_1(\alpha, u, v,\htheta,d_t ) - \frac{1}{N^2} F_1(\alpha l_g^2, u, v l_g ,0,d_t) \right)
\right \}\, ,
\ee
with:
\be
F_1( \alpha, u,v,\htheta,d) =  (\hc \alpha)^{d/2} \sum_{m,n\in \Z^d}   \exp \left \{- \pi  \hc \alpha  m^2 - \frac{\pi}{\hc u}
(n-\htheta  \tilde \epsilon m)^2  + 2 \pi i \frac{v}{u}  \, m n   \right\}
\, .
\ee
It is convenient, in order to analyze the infinite volume limit, to look at the expressions resulting after Poisson resummation in $m$ for both the $\htheta$-dependent and $\htheta$-independent parts. For the latter, Poisson resummation yields:
\be
F_1( \alpha, u,v,\htheta=0,d) =  \sum_{m,n\in \Z^d}   \exp \left \{- \frac{\pi }{ \hc \alpha} m^2 - \frac{\pi s}{\hc \alpha u} n^2
 + \frac{2 \pi v}{\hc \alpha u}  \, m n   \right\}
\, .
\ee
In the $\htheta$-dependent case, on the other hand, we begin by rewriting $m= \hat m l_g + m^c$, with the components of $m^c_\mu$ taking values in the intervals $[-l_g/2, l_g/2)$ or $[-(l_g-1) /2, (l_g-1)/2]$ when $l_g$ is respectively even or odd. Poisson resummation is then performed with respect to $\hat m$ only, leading to:
\begin{align}
F_1( \alpha, u,v,\htheta,d_t) =  \frac{1}{N^2} \sum_{m,n\in \Z^{d_t}} \sum_{m^c}  \exp & \left \{ - \frac{\pi }{ \hc \alpha l_g^2} m^2  
- \frac{\pi s }{ \hc \alpha u}  (n-\chi)^2  +
\frac{2\pi v}{ \hc \alpha u l_g}  m (n-\chi) \right  .  
\\
 &\left .  \, + \, \,  i \, \frac{2 \pi }{l_g} m m^c \right \} ,    
\nonumber
\end{align}
where we introduced a $d_t$-vector $\chi$ whose components are given by $\chi_\mu = ||\htheta \tilde \epsilon m^c_\mu||$, the symbol $||x||$ denoting the distance from $x$ to the nearest integer. Introducing $\chi_\mu = n^c_\mu / l_g$ and inverting the relation between $m^c$ and $n^c$ to write $m^c = k \epsilon n^c \, ($mod $l_g)$, we obtain:
\begin{align}
F_1( \alpha, u,v,\htheta,d_t) =  \frac{1}{N^2} \sum_{m,n\in \Z^{d_t}} \sum_{n^c} \exp & \left \{  - \frac{\pi }{ \hc \alpha u l_g^2} \Big (u m^2 + s (n l_g -n^c) ^2 - 2 v m (n l_g-n^c) \Big ) \right  .  
\\
 &\left .  \, + \, \,  i \, \frac{ 2 \pi k }{l_g} m \epsilon n^c \right \} .
\nonumber
\end{align}

The two terms entering $ \hat H(s,u,v,\htheta)$ and $\hat H(s,u,v,0)$ can be rewritten matricially. Recalling the expressions of $A_0$ and $B$ from eqs.~\eqref{eq:a0def} and~\eqref{eq:bdef} we have, in terms of Siegel theta functions, and particularizing to the case of $d_t=2$:
\begin{align}
& \hat H(s,u,v,0) = \Theta^2\left(0\Big |iA_0\left(\frac{1}{\hc \alpha},\frac{s}{\hc \alpha u},\frac{v}{\hc \alpha u}\right)\right)
\label{eq:HH0}
\\
&\left \{   \Theta^2\left(0\Big |iA_0\left(\frac{1}{\hc \alpha},\frac{s}{\hc \alpha u},\frac{v}{\hc \alpha u}\right)\right)
 - \frac{1}{N^2}
\Theta^2\left(0\Big |i A_0\left(\frac{1}{\hc N^2 \alpha},\frac{s}{\hc \alpha u },\frac{v}{\hc N \alpha u}\right)\right) \right\}
\, , 
\nonumber
\end{align}
and:
\begin{align}
& \hat H(s,u,v,\htheta) = \frac{1}{N^2} \Theta^2\left(0\Big |iA_0\left(\frac{1}{\hc \alpha},\frac{s}{\hc \alpha u},\frac{v}{\hc \alpha u}\right)\right)
\label{eq:HHt}
\\
&\left \{ {\rm Re} \Theta\left(0\Big |i B\left(\frac{1}{\hc N^2 \alpha},\frac{s}{\hc N^2 \alpha u },\frac{v}{\hc N^2 \alpha u}, \frac{k}{N}\right)\right)
 -
\Theta^2\left(0\Big |i A_0\left(\frac{1}{\hc N^2 \alpha},\frac{s}{\hc \alpha u },\frac{v}{\hc N \alpha u}\right)\right) \right\}
\, .
\nonumber 
\end{align}

We will now split the original integral into two pieces, setting $\htheta = 0$ in one part to confine all of the $\htheta$ dependence to the other one. As we want both of them to be well behaved both in the IR and in the UV, it will be convenient to first isolate the terms corresponding to zero-modes at each step of the calculation, both before and after Poisson resummation. In the original definition of $\hat H$, given by eq.~\eqref{eq:Hdef}, the terms with $m=0$ were already subtracted, so we simply need to take away the terms corresponding to $n=0$. Doing so leads to:
\begin{align}
\tI &= \frac{4}{3 \cA (2 \hc)} \int \left \{ (u \alpha)^{-2} \left(\hat H(s,u,v,0) -  \cA(2\hc ) \right) - \frac{\hc^2}{s^2}  \cA (\hc s)  \right \} \\
  &- \frac{4}{3 \cA (2 \hc)} \int \left \{ (u \alpha)^{-2} \hat H(s,u,v,\htheta) - \frac{\hc^2}{s^2} \cA (\hc s)  \right \} .
\nonumber
\end{align}
The analogous procedure after Poisson resummation, i.e. subtracting the expressions obtained setting $m=0$ and $n=0$ (separately) in eq.~\eqref{eq:HH0}, and adding back the $m=n=0$ one, yields:
\be
\hat H(s,u,v,0) = \hat H'(s,u,v,0) + \cA \left(\hc \alpha \right) + \frac{N^2-1}{N^2} \left \{\theta_3^4 \left(0,\frac{i s}{\hc \alpha u }\right) - 1 \right \} ,
\ee 
where $\hat H'$ denotes the resulting function after subtracting those zero modes. The same can be done for the term containing $\cA (\hc s)$, whose zero mode contribution is given by $-(1-1/N^2)c^2/s^2$.

The term $\hat H(s,u,v,\htheta)$ containing the $\htheta$ dependence requires a bit more work, but the idea is the same. We begin by rewriting the components of the 4-vector $n$ along the twisted directions as $n_\mu = \tilde n_\mu N + n_c$, with $n_c$ a 2-dimensional vector of integers taking values for $N$ even or odd in the respective intervals $[-N/2, N/2)$ or $[-(N-1) /2, (N-1)/2]$, and then subtract the terms corresponding to $n_\mu=0$ along periodic directions and $ \tilde n_\mu=0$ along the twisted ones. Subtracting the $m=0$ terms as well, and adding back once more the doubly subtracted ones, we end up with:
\begin{align}
\hat H(s,u,v,\htheta) &=  \hat H'(s,u,v,\htheta) - \cA \left(\frac{\hc \alpha u}{s} \right) + \frac{N^2-1}{N^2} \theta_3^4
\left(0, \frac{i s}{\hc \alpha u }\right) +  
\\
&\frac{1}{N^2} \sum_{n_c\ne 0} \exp\left \{ - \frac{\pi s n_c^2}{\hc N^2 \alpha u}\right\} {\rm Re}\left ( \theta_3^2\left (0,\frac{i}{\hc \alpha} \right) 
\prod_{\mu=0,1}  \theta_3\left (z_\mu,\frac{i}{\hc N^2 \alpha} \right) -1  \right ) ,
\nonumber
\end{align}
where $z_\mu =  \epsilon_{\mu\nu} n_{c\nu}  k/N + i n_{c \mu} v / (\hc N^2 \alpha u)$.

We may then rewrite each of the integrals contributing to $\cC_1$ as the the sum of two components $I=I_{TI}+I_{TD}$, the latter containing all of the $\htheta$ dependence:
\begin{align}
&I_{TI} &= I_{TI}^{(0)} + \frac{4}{3 \cA (2 \hc)} \left \{ \int (u \alpha)^{-2}  \hat H'(s,u,v,0)  + \int \frac{\hc^2}{s^2} \left ( 1 -\frac{1}{N^2} - \cA (\hc s)  \right ) \right \} , \\
&I_{TD} &= I_{TD}^{(0)}-  \frac{4}{3 \cA (2 \hc)} \left \{ \int (u \alpha)^{-2}  \hat H'(s,u,v,\htheta) + \int \frac{\hc^2}{s^2} \left ( 1 -\frac{1}{N^2} - \cA (\hc s)  \right ) \right \} , 
\end{align}
where:
\begin{align}
&I_{TI}^{(0)}= -\frac{4}{3 \cA (2 \hc)} \int (u \alpha)^{-2}  \left (  \cA (2 \hc) -  \cA (\hc\alpha ) - \cA (\hc\alpha u /s ) + 1 - \frac{1}{N^2} \right )
\, , 
\label{eq:iapprox_pE}\\
&I_{TD}^{(0)}=-\frac{4}{3 N^2 \cA (2 \hc)}   \sum_{n_c \ne 0}  \int (u \alpha)^{-2}  e^{ - \frac{ \pi s n_c^2}{\hc N^2 \alpha u } }
{\rm Re}\left \{ \theta_3^2\left (0, \frac{i}{\hc \alpha}\right) \prod_{\mu}
\theta_3 \left( z_\mu , \frac{i}{\hc N^2 \alpha}\right )
 - 1 \right \} \, ,
\label{eq:iapprox_npE}
\end{align}
and with $n_c$ and $z_\mu$ as defined above.

From this expression, one can analyze the $\hc\rightarrow 0$ limit, whose approach is driven by two variables: $\hc \alpha$ and $\hc \alpha u /s$. In all contributing integrals but $I_8$ and $I_9$, one of the two variables vanishes for all of the integration range when taking such a limit. The first thing worth noting is the fact that zero modes have already been subtracted from all terms not included in $I_{TI}^{(0)}$ and $I_{TD}^{(0)}$, and hence the leading order in the $\hc \rightarrow 0$ limit for them will be proportional to:
\begin{align}
-\frac{4}{3 N^2 \cA (2 \hc)} \int (u \alpha)^{-2} \exp\left \{- \frac{\pi}{\hc N^2 \alpha} - \frac{\pi s }{\hc N^2 \alpha u }+ \cdots \right \}, 
\end{align}
which approaches zero at least exponentially in the $\hc N^2 \rightarrow 0$ limit, and goes, in the large $N$ limit taken keeping $\hc N^2$ constant, as $1/N^2$. 
In most cases, the leading contribution in the $\hc\rightarrow 0$ limit is hence given by $I_{TI}^{(0)}$ and $I_{TD}^{(0)}$.

The simplest cases are those of $\tI_1$, $\tI_2$ and $\tI_4$, for which both  $\hc \alpha$ and $\hc \alpha u /s$ tend to zero in the whole integral range. Starting from the expressions of $I_{TI}$ and $I_{TD}$, it is easy to derive the leading correction to the large volume limit. In the three cases it is given by:
\be
\frac{16}{3 (N^2-1)} \int (u\alpha)^{-2} \left \{ e^{-\frac{\pi}{2 \hc N^2}} -  e^{-\frac{\pi}{\hc \alpha N^2}} - e^{-\frac{\pi s }{\hc \alpha u N^2}}
\right \}
\ee
All three integrals can be analytically approximated with this, leading to:
\begin{align}
&\tI_1 \rightarrow \frac{1}{9(N^2-1)}  e^{-(cN)^{-2}}   \left ( 1 + 3 \gamma_E - 3 \log \left (3 c^2 N^2\right) -3 c^2 N^2 \right )
\, ,
\\
& \tI_2 \rightarrow \frac{2}{9(N^2-1)}  e^{-(cN)^{-2}}    \left ( 1 -6 c^2 N^2 \right )
\, ,
\\
& \tI_4 \rightarrow -\frac{1}{3(N^2-1)}  e^{-(cN)^{-2}}    \left ( 1 -  \gamma_E +  \log \left(9 c^2 N^2\right) - 3.544907702 \, c N + c^2 N^2 \right)
\, .
\end{align}

We will now consider the remaining integrals, looking first at the $\hc$ dependence of  $I_{TI}^{(0)}$ and $I_{TD}^{(0)}$. For $\tI_3$, $\tI_6$ and $\tI_{10}$, the variable going to zero in the $\hc\rightarrow 0$ limit is $\hc \alpha u /s$, and the leading dependence is given by:
\be
\frac{4}{3 \cA (2 \hc)} \int (u \alpha)^{-2}  \left \{ \cA (\hc\alpha) - 1 +\frac{1}{N^2}  \right \} ,
\ee  
whereas for $\tI_5$, $\tI_7$, $\tI_{11}$, and $\tI_{12}$ the variable going to zero is $\hc \alpha$, and we have instead:
\be
\frac{4}{3 \cA (2 \hc)} \int (u \alpha)^{-2}  \left \{  \cA \left(\frac{\hc\alpha u}{ s} \right)  - 1 +\frac{1}{N^2} \right \} .
\ee
To leading order all these integrals go to zero as $\sim c^2$, with a coefficient depending on $N$ that is identical in absolute value for all of them.

\begin{figure}[t]
\centering
\includegraphics[width=0.95\linewidth]{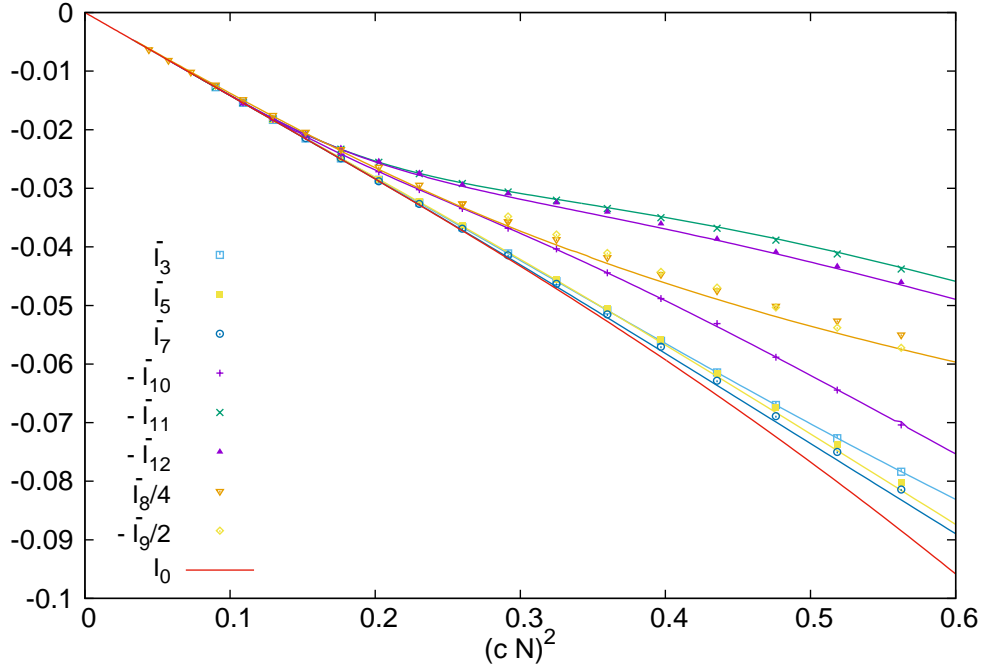}
\caption{The contribution to $\cC_1$ from the integrals $\tI_i$ with $i=5-10$, shown as a function $(c N)^2$. The continuous lines are obtained from
the approximate expression given by the sum of eq.~\eqref{eq:iapprox_pE} and ~\eqref{eq:iapprox_npE}. The red line represents $I_0$ as defined in
eq.~\eqref{eq:i30E}.}
\label{f:csq}
\end{figure}

We will take a look at $\tI_3$ as an illustrative example. The leading contribution in the $\hc\rightarrow 0$ limit for this integral is given by:
\be
-\frac{4}{3 \cA (2 \hc)} \int_0^\infty dz  (3+2z)^{-2} \left \{ 1 - \frac{1}{N^2} -  \cA (\hc (3+2z) /2 ) \right \},
\ee
which allows us to separate $\cA$ into two parts, one that depends on $N$ and another that does not:
\begin{align}
&\cA^{(1)}(x)= x^2 (\theta_3^4(0,i x) -1)\, ,  \\
&\cA^{(2)}(x)= x^2 (\theta_3^2(0,i x)\theta_3^2(0,i x N^2) -1)\, .
\end{align}
Rescaling $z$ to $z'=\hc z$ in the first expression and to $z'=\hc N^2 z$ in the second, we can decompose the integral into the difference of two pieces $\tI_3^{(1)}-\tI_3^{(2)}$, which in the $c N\rightarrow 0$ limit become:
\begin{align}
&\tI_3^{(1)} = \frac{\hc }{3 \cA(2 \hc)} \int_0^\infty d z \left \{ \theta_3^4\left (0,iz\right) -1- \frac{1}{z^2} \right \} , \\
&\tI_3^{(2)} = \frac{\hc }{3 N^2 \cA(2\hc)} \int_0^\infty d z \left \{ \theta_3^2\left (0,iz\right) \theta_3^2\left (0,iz/N^2\right)  -1- \frac{N^2}{z^2} \right \} .
\end{align}
The leading order result in the $cN \rightarrow 0$ limit is thus given by:
\be
I_0 = \frac{\pi (c N)^2}{ 6 N^2 \cA(2\hc) } \left(a_1 - \frac{1}{N^2} a_2(N)\right) + \cdots ,
\label{eq:i30E}
\ee
with $a_1= -1.76508480122121275$ and for instance $a_2(3) = 3.59085631503990722$.
One can show that all the other integrals are also proportional to $I_0$, with the proportionality coefficient being +1 for $i=5,6,7$ and -1 for $i=10,11,12$ respectively. The results for the case of the SU(3) gauge group are displayed on fig.~\ref{f:csq}, with the red line in the plot showing $I_0$ and the remaining continuous lines representing the contribution of $I_{TI}^{(0)}+I_{TD}^{(0)}$. The cases of $\tI_8$ and $\tI_9$ are shown in the plot as well, which also turn out to be proportional to $I_0$ with respective coefficients 4 and -2.

\bibliography{8_0_Refs}
\end{document}